\documentclass[a4paper,12pt,twoside,openany]{report}
\input epsf
\usepackage[dvips]{graphicx}
\usepackage{a4wide}
\usepackage{fancyhdr}
\usepackage[latin2]{inputenc}
\usepackage[T1]{fontenc}
\usepackage{latexsym}
\newcommand{\etmiss}{\mbox{\ensuremath{\not \hspace{-0.1cm}E_T}}}

\begin{document}
\title{Prospects for Observing an Invisibly Decaying Higgs Boson in $t\bar{t}H$ Production at the LHC}
\author{Maciej Malawski}
\date{May, 2004}

\thispagestyle{empty}
\begin{center}
\textsc{\Large Jagiellonian University} \\
\vspace*{2mm}
\textsc{Marian Smoluchowski} \\
\vspace*{2mm}
\textsc{\Large Institute of Physics } \\
\vspace{\baselineskip}
\textsf{\large Faculty of Physics, Astronomy and Applied Computer Science} \\
\vspace{.25\textheight}
\textbf{\huge 
Prospects for Observing an Invisibly Decaying Higgs Boson in $t\bar{t}H$ Production at the LHC
} \\
\vspace{2\baselineskip}
\textbf{\large Maciej Malawski} \\
\vspace{5\baselineskip}
Master of Science Thesis\\
Physics \\
\vfill
\begin{tabular}{l}
{\large \textsl{Supervisor:} Prof. Elżbieta Richter-Wąs} \\
\end{tabular} \\
\vspace{2\baselineskip}
Krak\'ow, May 2004 \\
\end{center}

\pagenumbering{roman}
\newpage\verb||\thispagestyle{empty}\newpage

\thispagestyle{empty}
\renewcommand\abstractname{Acknowledgements}
\begin{abstract}
I would like to express my thanks to my supervisor -- Prof. Elżbieta
Richter-Wąs -- 
for introducing me into the fascinating research area of particle physics
and for her invaluable help and advice during preparation of this thesis.

\noindent I also gratefully acknowledge the support of my wife, Krysia, who 
continuously motivated me to finish this work.

\vspace{5mm}
\hspace{3cm}Maciej Malawski
\end{abstract}
\newpage\verb||\thispagestyle{empty}
\newpage

\renewcommand\abstractname{Abstract}
\begin{abstract}
In this thesis we study the prospects for observing the invisibly decaying Higgs boson
in the associated $t\bar tH$ production at the LHC.
The results of the Monte Carlo simulations of signal
and background processes show that there is a possibility
of observing the statistically significant number of signal events
required for the discovery. Moreover, the analysis can be further improved
to reduce the number of false reconstructions of the $W$ boson.

The analysis of the $t \bar t H$ production is independent of the model in which the Higgs
boson decays into the invisible channel.
There are several possibilities for models where $H\to invisible$ can be of interest.
For this thesis, we have studied the simplest supersymmetric model,
called mSUGRA.
The results of the scans of the mSUGRA model parameter space
show that the regions, where the branching ratio of the lightest neutral Higgs boson to the lightest neutralino
pair is high, are excluded by current experimental constraints. The
$h \to b \bar b$ channel dominates, and the possibility for discovery in this channel
will not be suppressed by the invisible decays. 
This result does not disqualify invisible channel as possible signature in other models.

Chapter~\ref{chap:analysis} of this thesis is based on the published paper: 
B.P. Kersevan, \underline{M. Malawski}, E.  Richter-Wąs: {\em Prospects for observing an invisibly decaying Higgs boson in the $t\bar t H$
 production at the LHC}, The European Physical Journal C - Particles and Fields, 2003, vol. 29, no. 4, pp. 541 - 548.

\end{abstract}

\newpage\verb||\thispagestyle{empty}
\newpage
\thispagestyle{empty}
\tableofcontents
\thispagestyle{empty}
\listoffigures
\thispagestyle{empty}
\listoftables
\thispagestyle{empty}
\newpage\verb||\thispagestyle{empty}
\newpage
\chapter{Introduction}

\pagenumbering{arabic}

The Higgs boson is a very important element of the modern
theory of elementary particles and their interactions, called the Standard Model.
By means of {\em spontaneous symmetry breaking} the {\em Higgs mechanism}
 is used to introduce mass to other particles.
The Higgs boson remains the last undiscovered particle predicted by the Standard Model.
The most important large-scale experiments of these days, at the Large Hadron Collider at CERN
and at Tevatron at Fermilab aim at the discovery of this particle.

Worldwide collaborations of scientists are studying the possible production and decay channels,
where the Higgs boson may be observed. There are theories, in which the Higgs
boson can decay into the {\em invisible} channel. This can occur when the Higgs 
decays into particles which interact very weakly with the matter, so they cannot be observed
in the detector. There are various models that may lead to such invisible decays.
These models include the decay into the lightest neutralinos in the supersymmetry models,
or into neutrinos in the models of the neutrino mass generation,
such as extra dimensions, TeV-scale gravity, Majorana models or 4th generation lepton~\cite{Godbole03}.

The associated production of the Higgs boson and a $t\bar t$-quark pair 
can be an interesting channel leading to the observation of the invisible Higgs boson.
The top quark production is a process with a very characteristic topology, so it can be used to
effectively filter the events in which Higgs production takes place.
One of the objectives of this thesis was to study the possibility 
of observing the Higgs boson in this channel using 
simulations the Atlas detector at LHC.

Supersymmetry (SUSY) is considered by many theorists
as the most probable extension of the Standard Model. 
SUSY assumes a symmetry between fermions and bosons, introducing
superpartners for each of the known elementary particles.
The Minimal Supersymmetric extension of the Standard Model (MSSM)
predicts the existence of the Lightest Supersymmetric Particle (LSP),
which does not decay into normal particles and is a candidate for dark matter.
A Higgs boson, decaying into a pair of LSPs, can give an invisible signature.

In order to obtain a more realistic estimation of the $t\bar t H$ process analysis results,
it is necessary to discern the cross-section of the $t\bar t H$ production and
the branching ratio of the Higgs to the invisible channel.
SUSY models allow for such predictions. In this thesis, we describe
the results of the scans of the mSUGRA model parameter space.
The objectives of the scans were to find such regions in the parameter space, 
where the branching ratio of the Higgs into the invisible channel is large enough
to make the results of the analysis statistically significant.
The production process was studied only for the lightest Higgs boson
in the so-called decoupling limit, i.e. in the region where its
coupling to fermions and bosons is almost equal to that of the Standard Model Higgs boson
of the same mass.

The thesis is organized as follows: Chapter~\ref{chap:higgs} provides a
theoretical introduction to the Standard Model Higgs boson 
and gives an overview of the current
and planned experiments for Higgs boson discovery.
Chapter~\ref{chap:susy} provides a brief introduction to supersymmetry and 
the specific mSUGRA model that can lead to an invisible Higgs decay.
In Chapter~\ref{chap:analysis} we describe the analysis of the $t\bar t H, H\to invisible$ process
based on Monte Carlo simulations of the ATLAS detector at LHC.
Chapter~\ref{chap:scans} presents the results of the scans of the mSUGRA 
model parameter space, where the invisible decay channel is possible.
The conclusions are summarised in Chapter~\ref{chap:conclusions}.

\chapter{The Higgs Boson}
\label{chap:higgs}
\section{The Standard Model and the Higgs Theory}

The Standard Model (SM for short), is currently regarded as a successful theory describing
elementary particles and their interactions.
SM is based on the quantum field theory. It assumes the existence of 
fermion fields representing matter and gauge fields that carry interactions.
Matter is composed of three generations of quarks and leptons. 
For the spinor fields $\psi$, we have the Dirac Lagrangian:

\begin{equation}
\label{eq-ldirac}
L = \bar \psi (i\gamma^\mu\partial_\mu - m)\psi.
\end{equation}
>From this Lagrangian we can derive the Dirac equation, which is the equation of motion
of free relativistic particles and antiparticles (fermions).
In order to introduce interactions into the model, we postulate the gauge invariance of the fields.
The Lagrangian should be invariant under the $U(1)$ gauge transformation:

\begin{equation}
\psi \to e^{-i\alpha(x)}\psi.
\end{equation}
For the Lagrangian~(\ref{eq-ldirac}) to be invariant, we need to introduce
a new vector field $A_\mu$ and derive a new Lagrangian:

\begin{equation}
\label{eq-lgauge}
L_D = L - e \bar \psi \gamma^\mu A_\mu \psi ,
\end{equation}
where the transformation law for the $A_\mu$ vector field is:

\begin{equation}
A_\mu \to A'_\mu = A_\mu + \frac{1}{e}\partial_\mu\alpha(x) .
\end{equation}
To the Lagrangian~(\ref{eq-lgauge}) we also add a term that yields the equations 
for gauge fields. For the electromagnetic field, the term is:

\begin{equation}
L_A = - \frac{1}{4}F_{\mu\nu}F^{\mu\nu}, F_{\mu\nu} = \partial_\mu A_\nu - \partial_\nu A_\mu.
\end{equation}
Such a Lagrangian can provide the Maxwell equations for electromagnetic fields 
and the Dirac equations for fermions:

\begin{equation}
 \Box A^\nu = e \bar \psi \gamma^\nu \psi = j^\nu;\quad
 \gamma^\mu(i\partial_\mu - e A_\mu)\psi - m\psi = 0.
\end{equation}
In the case of a multiplet of $N$ particles, we have a more general Dirac Lagrangian:

\begin{equation}
L = \bar \Psi (i\gamma^\mu\partial_\mu - M)\Psi ,
\end{equation}
where $\Psi$ is an $N-$component vector. When we require the gauge invariance of
such a theory, the transformation has the following form:

\begin{equation}
\Psi' = G \Psi,
\end{equation}
where $G$ is the $N \times N$ unitary matrix, $det G = 1$. These 
matrices form the group $SU(N)$. The $SU(N)$ invariance
requires the introduction of $N^2-1$ vector fields $G^n_\mu$ and modification of the Lagrangian:

\begin{equation}
L_D = L - g\bar\Psi\gamma^\mu\sum_n T^n G^n_\mu \Psi,
\end{equation}
where $T^n$ are generators of the $SU(N)$ group. By introducing the operators: 

\begin{equation}
\hat G_\mu = \sum_n G^n_\mu T^n,\quad \hat G_{\mu\nu} = \partial_\mu \hat G_\nu - \partial_\nu \hat G_\mu + ig[\hat G_\mu, \hat G_\nu ],
\end{equation}
we can construct the Lagrangian for the vector fields:

\begin{equation}
L_G = - \frac{1}{2}Tr(\hat G_{\mu\nu} \hat G^{\mu\nu})
\end{equation}
The full Lagrangian of the $SU(N)$ theory has the following form:

\begin{equation}
L_{YM} = \bar \Psi (\gamma^\mu[i\partial_\mu  - g\hat G_\mu] - M) \Psi - \frac{1}{2}Tr(\hat G_{\mu\nu} \hat G^{\mu\nu})
\end{equation}

To describe the electroweak interactions for one generation of leptons or quarks,
we need the $SU(2)_L\times U(1)$ group. The doublet $\Psi$ consists of the (lepton, neutrino)
pair or (up-, down-)type quark pair respectively. The left-handed components $\Psi_L \equiv L$
of these spinors transform as the fundamental representation of the $SU(2)$ group,
while the right-handed components $\Psi_R \equiv R$ are singlets with respect to this group. 
All members of the doublet transform under the $U(1)$ group. 
Consequently, we need four vector fields, $B_\mu$ corresponding to $U(1)$ and  $B^k_\mu, (k=1,...,3)$ for $SU(2)$.
These fields give the fermion Lagrangian:
\begin{equation}
\label{eq-lfermion}
\begin{array}{lcl}
L_F &=& i\bar L\gamma^\mu\partial_\mu L - \bar L \gamma^\mu(\frac{1}{2}g'Y_LB_\mu + gB^k_\mu T^k)L\\
 & +& i\bar\psi^1_R\gamma^\mu\partial_\mu \psi^1_R - \bar \psi^1_R \gamma^\mu\frac{1}{2}g'Y_{1R}B_\mu\psi^1_R\\
 & +& i\bar\psi^2_R\gamma^\mu\partial_\mu \psi^2_R - \bar \psi^2_R \gamma^\mu\frac{1}{2}g'Y_{2R}B_\mu\psi^2_R .\\
\end{array}
\end{equation}
In this Lagrangian, the weak hypercharge is denoted by $Y$.
The Lagrangian for vector fields has the following form:

\begin{equation}
\label{eq-lvector}
L_V = - \frac{1}{4}B_{\mu\nu}B_{\mu\nu} - \frac{1}{4}\sum_k B^k_{\mu\nu}B^k_{\mu\nu}.
\end{equation}
The fermion Lagrangian (\ref{eq-lfermion}) cannot contain the mass term $\bar \Psi \Psi$ which
is not invariant with respect to the $SU(2)$ group, since the left and right components transform differently, and:

\begin{equation}
\bar \Psi \Psi = \bar \Psi_R\Psi_L + \bar\Psi_L\Psi_R.
\end{equation} 
Moreover, $B^k_\mu$ fields are also massless, 
since we do not obtain the linear mass terms in motion equations from the Lagrangian (\ref{eq-lvector})..

The model with all particles being massless is unsatisfactory, since we observe massive 
particles in the experiments. The solution to this problem is the {\em Higgs mechanism}.
We add to our model one more $SU(2)$ doublet, consisting of two complex scalar fields:

\begin{equation}
\label{eq-higgs-sm}
H=\left[\begin{array}{c} h_1 \\ h_2 \end{array}\right],
\end{equation}
called the Higgs field.  
The Lagrangian for these fields has to be gauge-invariant under $U(1)$ and $SU(2)$. We can choose
such a gauge in which the Higgs field has the form:

\begin{equation}
H=\left[\begin{array}{c} 0 \\ h \end{array}\right],
\end{equation}
where $h$ is the real function. This means that there is only one physical degree of freedom,
so the Higgs field can give one neutral scalar particle. 
Very important in the construction of the Higgs Lagrangian 
is the substitution of the mass term by the Higgs potential. We postulate:

\begin{equation}
L_H = L_H^{free} - V = \partial_\mu H^\dag \partial_\mu H - \mu^2H^\dag H - \lambda[H^\dag H]^2.
\end{equation}
When $\mu^2<0$ the Higgs potential has a non-trivial minimum as shown in Fig.~\ref{fig:potential}.
The minimum is for:

\begin{equation}
H^\dag H = - \frac{\mu^2}{2\lambda}.
\end{equation}
We can choose the phase of the solution as:

\begin{figure}[htb]
\begin{center}
\includegraphics[width=0.5\textwidth, height=0.3\textwidth]{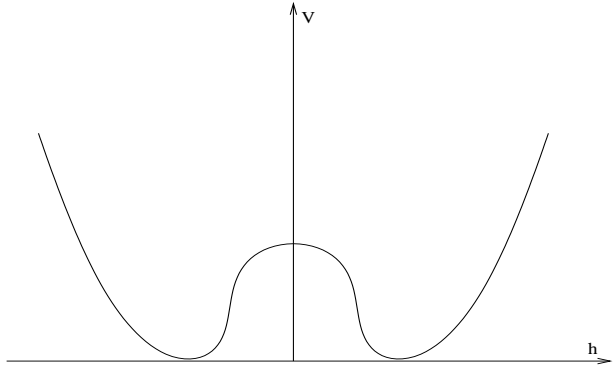}
\label{fig:potential}
\caption{Mexican-hat Higgs potential with a non-zero minimum}
\end{center}
\end{figure}

\begin{equation}
H = H_0 = \left[\begin{array}{c} 0 \\ v/\sqrt{2} \end{array}\right];\quad v^2 = - \frac{\mu^2}{\lambda}
\end{equation}
Such a solution breaks the gauge symmetry. This is called the spontaneous symmetry breaking, since 
the potential itself is gauge-invariant, whereas the vacuum state breaks the symmetry.
The Lagrangian containing interactions of the Higgs and the vector fields has the following form:

\begin{equation}
\label{eq-lew}
L = - \frac{1}{4}B_{\mu\nu}B_{\mu\nu} - \frac{1}{4}\sum_k B^k_{\mu\nu}B^k_{\mu\nu} 
+ \frac{v^2}{8}((g'B_\mu - gB^3_\mu) + g^2(B^1_\mu B^1_\mu + B^2_\mu B^2_\mu))
\end{equation}
This Lagrangian contains the terms bilinear in fields. We need to diagonalize this expression and derive
mass coefficients.
For components $B^1_\mu$ and $B^2_\mu$ the Lagrangian is already diagonal,
so we obtain the mass terms:

\begin{equation}
M^1 = M^2 = \frac{gv}{2}
\end{equation}
We can form combinations for particles with a definite charge:

\begin{equation}
W^\pm = \frac{B^1_\mu \pm B^2_\mu}{\sqrt 2}.
\end{equation}
These states form the charged $W$ bosons that carry 
the electroweak interactions.
We also have combinations that yield the neutral bosons:

\begin{equation}
Z_\mu = \frac{-g'B_\mu + g B^3_\mu}{\sqrt{g'^2 + g^2}}.
\end{equation}
The mass of these bosons is:

\begin{equation}
M_Z = \frac{v\sqrt{g'^2 + g^2}}{2}
\end{equation}
The remaining combination diagonalizing (\ref{eq-lew}) is 

\begin{equation}
A_\mu = \frac{g'B_\mu + g B^3_\mu}{\sqrt{g'^2 + g^2}}.                        
\end{equation}
This component is not present in the Lagrangian, so the mass coefficient must
vanish. The massless field represents the photon. 
The ratio between the masses of $W$ and $Z$ bosons defines the
Weinberg angle:

\begin{equation}
\frac{M_W}{M_Z} = \frac{g}{\sqrt{g'^2+g^2}} \equiv \cos \theta_W .
\end{equation}
To construct the Lagrangian for the Higgs field, we can rewrite it
by distinguishing the vacuum part from the dynamical part:

\begin{equation}
H(x) = H_0 + h(x),\quad h(x) \equiv \left[\begin{array}{c} 0 \\ h(x) \end{array}\right].
\end{equation}
The Lagrangian for this field can be written in the form:

\begin{equation}
L_H = \partial_\mu h\partial^\mu h + 2 \mu^2h^2 - 2\sqrt 2 \lambda v h^3 - \lambda h^4
+ \frac{g}{cos\theta_W}[vh + h^2][Z_\mu Z_\mu + 2 \cos\theta_W W^+_\mu W^-_\mu].
\end{equation}
This Lagrangian gives the coupling of the Higgs field to the vector bosons
and the Higgs mass:

\begin{equation}
m_H = -2\mu^2 = 2\lambda v.
\end{equation}
This means that the value of the $v$ parameter is not enough to determine the Higgs mass,
but it  also depends on $\lambda$, which is a free parameter of the Standard Model.

The Higgs mechanism also gives mass to fermions. This is done by introducing
the Yukawa couplings to the Lagrangian:

\begin{equation}
L_Y = -G_1[\bar L \tilde H \psi_{1R} + \bar\psi_{1R}\tilde H ^\dag L]
- G_2[\bar L H \psi_{2R} + \bar \psi_{2R}H^\dag L],
\end{equation}
where

\begin{equation}
\tilde H = [h_0 + h(x)]\left[\begin{array}{c} 1 \\ 0 \end{array}\right].
\end{equation}
For $h(x) \equiv 0$ we obtain

\begin{equation}
L_Y = -G_1h_0\bar\psi_1\psi_1 - G_2h_0\bar\psi_2\psi_2,
\end{equation}
giving us the fermion masses:

\begin{equation}
m_1 = G_1h_0; \quad m_2 = G_2h_0.
\end{equation}
Similarly, we obtain the couplings of the Higgs field to fermions
in the Lagrangian:

\begin{equation}
L_Y = -G_1h(x)\bar\psi_1\psi_1 - G_2h(x)\bar\psi_2\psi_2.
\end{equation}
>From these equations we can see that the coupling of the Higgs boson
to fermions is proportional to the fermion mass. 
As a consequence, we can predict that the Higgs boson
will mostly decay into the heaviest lepton or quark pairs.

\section{Current Experimental Results}

The Standard Model predictions have passed all experimental tests conducted so far.
The electroweak gauge bosons $W$ and $Z$ were detected at the LEP experiment
at CERN and the $t$~quark was observed at the Tevatron.
Only the Higgs boson remains as the last undiscovered elementary particle of the Standard Model.
Discovery of the Higgs boson will enable verification of theoretical foundations 
of the Standard Model.

\begin{figure}[htb]
\begin{center}
\includegraphics[width=0.25\textwidth]{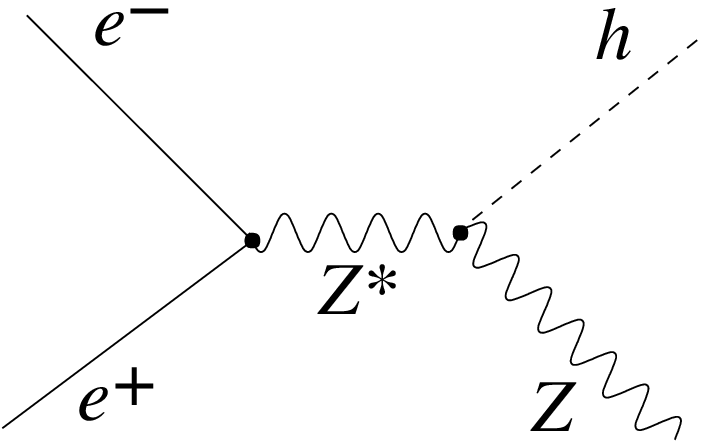}\hspace{1cm}
\includegraphics[width=0.25\textwidth]{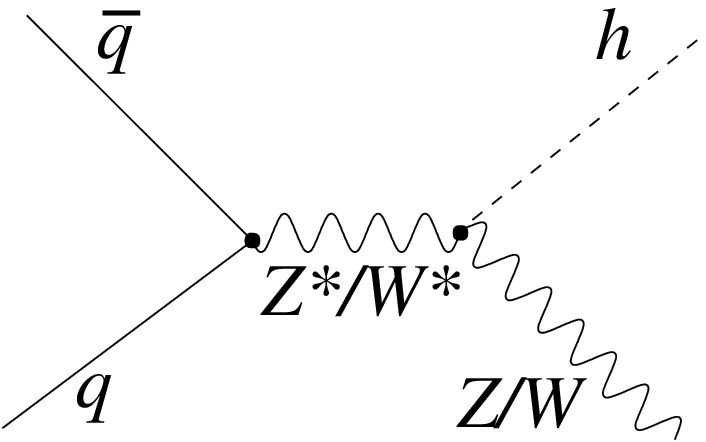}\hspace{1cm}
\includegraphics[width=0.25\textwidth]{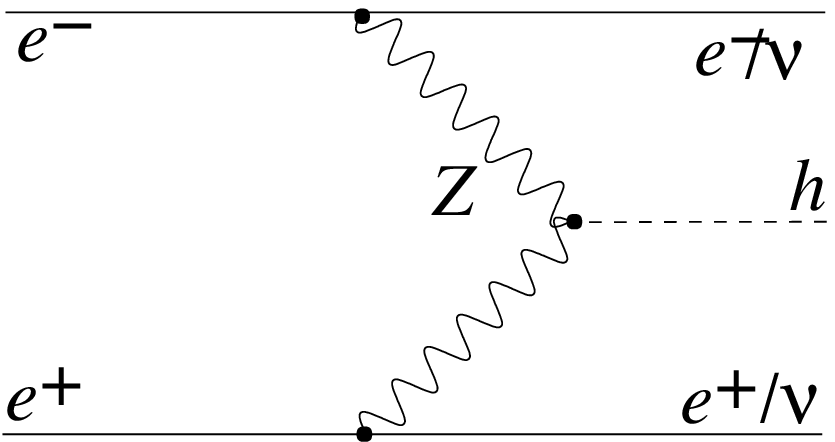}
\label{fig:higgsprod}
\caption{Processes of Higgs production}
\end{center}
\end{figure}

The possibilities of observing the Higgs particle
both in $e^{+}e^{-}$ and in hadron colliders have been extensively analyzed.
Since the coupling of the Higgs to other particles is proportional to their masses,
the Higgs production and decay often involve the heavy quarks $t$ or $b$.
Sample processes of Higgs production, such as {\em Higgs-strahlung} and 
weak boson fusion are shown in Fig~\ref{fig:higgsprod}.

\begin{figure}[ht]
\begin{center}
\includegraphics[width=0.45\textwidth,angle=-90]{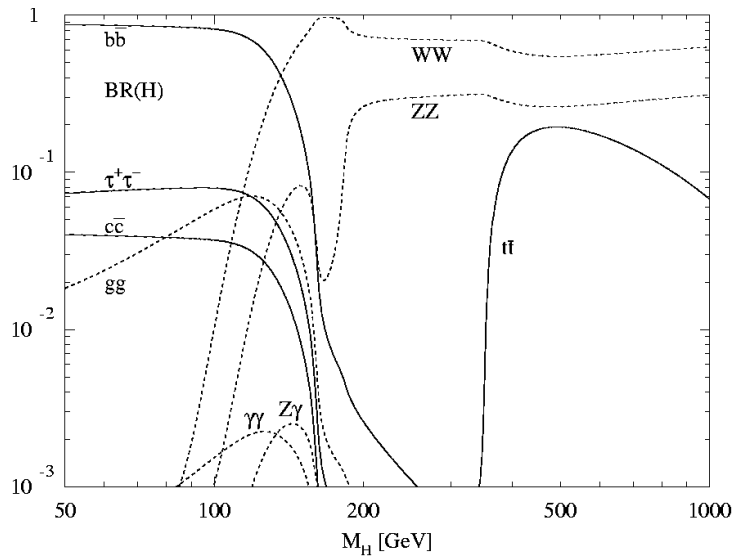}
\caption{The branching ratios of the Standard Model Higgs boson~\cite{Djouadi:1997yw}.
\label{fig:sm-higgs-br}
}
\end{center}
\end{figure}

The branching ratios of the Higgs boson to the others particles
depend on the Higgs mass. Fig.~\ref{fig:sm-higgs-br}
shows the branching ratios for various possible decay channels.

\begin{figure}[ht]
\begin{center}
\includegraphics[width=0.55\textwidth]{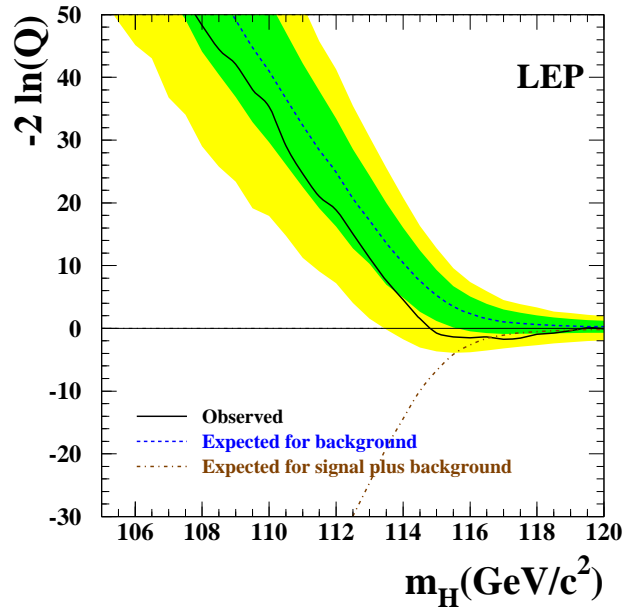}
\caption{The results of direct search for the SM Higgs boson at LEP
\cite{lep-final}.
\label{fig:lep-higgs-final}
}
\end{center}
\end{figure}

The results gathered at LEP from direct search for the Higgs boson
and from the precision electroweak measurements provide some 
approximate limits to the Higgs boson mass.
Direct search at LEP has excluded the existence of the Higgs boson
with a mass $m_H < 114.4$~GeV, at a 95\% confidence level.
The Fig~\ref{fig:lep-higgs-final}
shows the expected and observed behavior of the statistic 
defined in \cite{lep-final}.

\begin{figure}[ht]
\begin{center}
\includegraphics[width=0.5\textwidth]{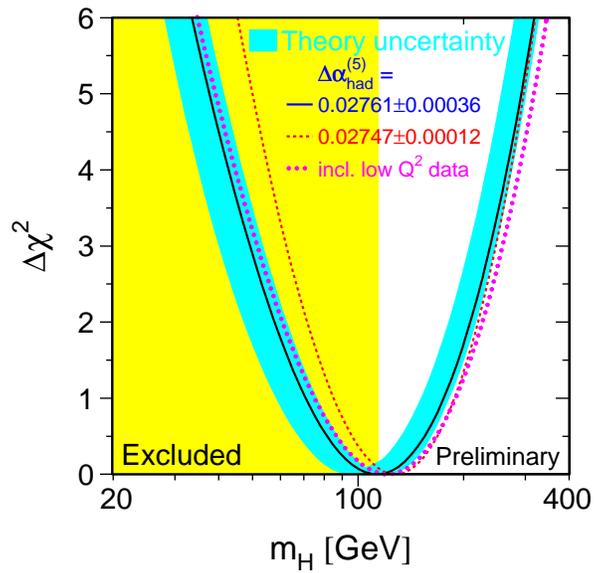}
\caption[The fit to electroweak measurements at LEP]{The fit to electroweak measurements at LEP, setting bounds
on the Higgs boson mass~\cite{lep-ew-final}
\label{fig:blueband}
}
\end{center}
\end{figure}

The precise measurements of the electroweak boson masses and couplings
can also give an estimation of the Higgs boson mass. The
combined results from the LEP and Tevatron experiments are shown 
in Fig.~\ref{fig:blueband}. They suggest the preferred Higgs boson mass
to be: $m_H = 113 (+ 62 - 42)$~GeV at a 68\% confidence level.
They also set the upper limit to be about 237 GeV at a 95\% confidence level.

\section{LHC Discovery Potential}

\begin{figure}[ht]
\begin{center}
\includegraphics[width=0.5\textwidth]{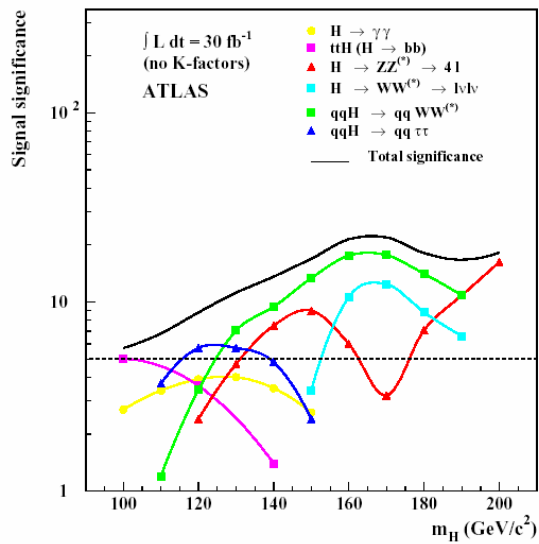}
\caption{ATLAS discovery potential for the SM Higgs boson~\cite{atlas-potential}
\label{fig:atlas-potential}
}
\end{center}
\end{figure}

The LHC experiment will search for the SM Higgs boson
in collisions of proton-proton pairs.
The ATLAS and CMS experiments are preparing for 
possible discovery channels. Fig.~\ref{fig:atlas-potential}
shows the ATLAS discovery potential for the SM Higgs 
boson for the integrated luminosity of 30 fb$^-1$~\cite{atlas-potential}. 
It can be seen that the total 
estimated significance is higher than the $5\sigma$ discovery threshold for
the Higgs mass range between 100 and 200 GeV.

\chapter{Supersymmetry}
\label{chap:susy}
\section{Introduction to Supersymmetric Theories}

In spite of their success
and compatibility with experimental results, both the Standard Model and the Higgs mechanism 
outlined in Chapter~\ref{chap:higgs} 
remain unsatisfactory as theories~\cite{ellis}.
One of the problems that the Standard Model does not solve 
is the {\em hierarchy problem}. This is a problem of the 
sensitivity of the Higgs mass to quantum corrections
from the particles coupling to the Higgs boson. It so happens
that the corrections from bosons and from fermions have opposite signs.
Thus, the corrections would cancel each other out if each of the known fermions had a bosonic partner 
of the same mass and each boson had a fermionic partner. Such a symmetry
is called supersymmetry~\cite{martin}.

Supersymmetry transforms boson states into fermions, and vice versa. This can be expressed as follows:

\begin{equation}
Q|Boson\rangle = |Fermion\rangle; \quad Q|Fermion\rangle = |Boson\rangle
\end{equation}
The supersymmetry generator $Q$ is an anticommuting spinor that satisfies 
the following algebra:

\begin{eqnarray}
 \{ Q,Q^\dag \} & = & P^\mu \\
 \{ Q,Q \} & = & \{ Q^\dag, Q^\dag \} = 0 \\
  \left[ P^\mu , Q\right]   & = &  \left[P^\mu , Q^\dag\right]  = 0
\end{eqnarray}
Supersymmetric particles can form multiplets, consisting of bosonic and fermionic
states, called supermultiplets. All particles and their superpartners
from the same supermultiplets have the same masses, since the $P^2$
operator commutes with the $Q, Q^\dag$. The supersymmetry generators 
commute also with the generators of gauge transformations. Hence, 
particles in the same supermultiplet have the same electric charge, weak isospin
and color degrees of freedom. In each supermultiplet, the number of fermionic degrees 
of freedom $n_F$ is equal to the number of bosonic degrees of freedom $n_B$:

\begin{equation}
n_F = n_B
\end{equation}

The simplest possibility for a supermultiplet includes a single Weyl fermion with two helicity
states and a complex scalar field with two degrees of freedom. Such a combination is
called the {\em chiral} or {\em matter} supermultiplet. The next simplest possibility
is the supermultiplet with a massless spin-1 vector boson, which has two degrees of freedom.
Its superpartner is a massless spin-1/2 Weyl fermion. Such a combination is 
called the vector supermultiplet.

\section{Minimal Supersymmetric Standard Model}

The Standard Model particles and their superpartners are presented 
in Table~\ref{tab-spartners}. The table shows the states before
the electroweak symmetry is broken, i.e. the $W$ and $B$ bosons correspond
to $B^k_\mu$ and $B_\mu$ fields from Lagrangian~(\ref{eq-lvector})
and their partners are called winos and binos. 
After breaking this symmetry,
new states appear: zino $\tilde Z$ and photino $\tilde \gamma$.
All these particles and their interactions form the Minimal Supersymmetric extension 
to the Standard Model (MSSM).

In Table~\ref{tab-spartners} we can see two Higgs doublets: $H_u, H_d$.
In the Standard Model, one complex scalar Higgs doublet
was enough (see eq.~(\ref{eq-higgs-sm})). Now, however, two complex doublets
are required to give mass to the particles. 

\begin{table}[htb]
\begin{center}
\caption[The Standard Model particles and their
superpartners]{The Standard Model particles and their 
superpartners before electroweak symmetry breaking
\label{tab-spartners}
}
\begin{tabular}{cccccc}\\ \hline\noalign{\smallskip}
Particle 	&		&Spin	&Spartner	&				&spin\\ \noalign{\smallskip}\hline\noalign{\smallskip}
quarks: 	&$u, d$ 	& 1/2 	& squarks: 	&$\tilde u, \tilde d$		& 0\\
lepton: 	&$l$ 		& 1/2	& sleptons: 	&$\tilde l$			& 0\\
Higgs: 		&$H_u, H_d$	& 0	& higgsino: 	&$\tilde H_u, \tilde H_d$	& 1/2\\
gluon: 		&$g$		& 0	& gluino: 	&$\tilde g$			& 1/2\\
W bosons: 	&$W$		& 0 	& winos: 	&$\tilde W$			& 1/2\\
B bosons: 	&$B$		& 0	& binos: 	&$\tilde B$			& 1/2 \\\hline
\end{tabular} 
\end{center} 
\end{table}

\noindent
The free Lagrangian of the supersymmetric model can be written in the following form:

\begin{equation}
L_{\mbox{free}} = -\partial ^\mu \phi^{*i}\partial_\mu \phi_i - i\psi^{\dag i}\bar\sigma^\mu\partial\mu\psi_i + F^{*i}F_i,
\end{equation}
where $\phi$ are the scalar fields, $\psi$ are the Weyl spinors from the supermultiplet
and $F$ is the auxiliary scalar field, required to fill the remaining scalar degrees of freedom.

The interaction can be introduced into the theory by adding the Lagrangian

\begin{equation}
L_{\mbox{int}} = - \frac{1}{2}W^{ij}\psi_i\psi_j + W^i F_i + c.c.,
\end{equation}
where $W^{ij}$ and $W^i$ are the functions of the bosonic fields $\phi_i$.
These terms can be expressed using the superpotential $W$:

\begin{equation}
W = \frac{1}{2}M^{ij}\phi_i\phi_j + \frac{1}{6}y^{ijk}\phi_i\phi_j\phi_k,
\end{equation}
and relations:

\begin{equation}
W^{ij} = \frac{\delta^2}{\delta\phi_i\delta\phi_j}W;\quad W^i = \frac{\delta W}{\delta\phi_i},
\end{equation}
where $M^{ij}$ is the mass matrix for the fermion fields, and $y^{ijk}$ is a Yukawa
coupling between the two fermions $\psi_i\psi_j$ and the scalar $\phi_k$.

In the case of the MSSM, the superpotential has the form:

\begin{equation}
W_{\mbox{MSSM}} = \bar u y_u Q H_u - \bar d y_d Q H_d - \bar e y_e L H_d + \mu H_u H_d.
\end{equation}
In this equation we use the superfield notation, i.e. $H_u, H_d, \bar u$, etc. denote the 
chiral supermultiplets, as in Tab.~\ref{tab-spartners}. The $\mu$ term corresponds to the Higgs mass in the Standard Model.

\section{Models of Supersymmetry Breaking}

The experimental results show that supersymmetry has to be broken,
since no sparticles with masses equal to their Standard Model partners have been discovered.
Despite the symmetry breaking, we would like to preserve the supersymmetric relations
between superpartners in the high energy sector of the theory, in order to keep
the solution to the hierarchy problem valid. Such an approach is called the
soft supersymmetry breaking.
This can be expressed by an effective Lagrangian of the MSSM:

\begin{equation}
\label{eq-lmssm}
\begin{array}{ccl}
L^{MSSM}_{soft} & = & -\frac{1}{2}(M_3\tilde g \tilde g + M_2 \tilde W \tilde W + M_1 \tilde B \tilde B) + c.c.\\
 & - & (\tilde{ \bar u} a_u \tilde Q H_u - \tilde{ \bar d} a_d \tilde Q H_d - \tilde{ \bar e} a_e \tilde L H_d) + c.c.\\
 & - & \tilde Q ^\dag m^2_q \tilde Q - \tilde L^\dag m^2_L \tilde L - \tilde{ \bar u} m^2_{\bar u} \tilde{ \bar u}^\dag - \tilde{ \bar d} m^2_{\bar d} \tilde{ \bar d}^\dag - \tilde{ \bar e} m^2_{\bar e} \tilde{ \bar e}^\dag\\
 & - & m^2_{H_u}H^*_u H_u - m^2_{H_d}H^*_d H_d - b H_u H_d + c.c.).
\end{array}
\end{equation}
We can see that in eq. (\ref{eq-lmssm}) we have the $M_1, M_2$ and $M_3$ gaugino mass terms, $a_u, a_d, a_e$ are 
the matrices of Yukawa couplings, and the $m^2_i$ terms are mass matrices of fermion families.
Together with the Higgs sector parameters, it is necessary to have 105 free parameters 
in the MSSM.
The experimental results that restrict the flavor-changing neutral currents and CP violation
suggest that the squark and slepton mass matrices should be diagonal. This organizing principle helps reduce the number of
parameters. 

Such universality can be achieved by 
assuming that, at high energy scale, there exists some simple Lagrangian
in which there are only a few common mass and coupling parameters.
After applying the renormalization group equations to the masses and couplings
to compute their values in the electroweak scale, we can obtain the full spectrum
of the MSSM parameters. 
Such unification of couplings is in agreement with the GUT theories, 
which predict unification at the mass scale of $10^{16}$GeV.
Since the high-energy scale theory cannot be observed directly,
it is called the hidden sector, while the MSSM is called the visible sector.

There is more than one model that can lead to such supersymmetry breaking.
All these models can be divided with respect to the type of interaction between the hidden sector and the visible one.
In gravity-mediated sypersymmetry breaking one assumes the supergravity 
Lagrangian in the hidden sector and the gravitational interactions lead to
supersymmetry breaking. On the other hand, in the gauge-mediated supersymmetry breaking models,
one assumes that there exist some additional chiral supermultiplets, called messengers,
which couple to the fields of the hidden sector and also to the MSSM fermions through gauge interactions.

An especially interesting model, called mSUGRA, is the minimal version
of gravity-mediated supersymmetry breaking models. It assumes unifications
of all fermion masses into $m_{1/2}$, all boson masses into $m_0$, and one common 
coupling constant $A_0$. To complete the model, it is necessary to add the
$\tan \beta$ and sign~$\mu$ parameters describing the Higgs sector. Thus, there are only 5 parameters:

\begin{equation}
\label{eq-msugra}
m_0, m_{1/2}, A_0, \tan \beta, \mbox{sign} \mu,
\end{equation}
that can give the full MSSM spectrum after applying renormalization group equations
to the electroweak scale.

The mass spectrum of the MSSM is subject to the electroweak symmetry breaking,
similar to that of the Standard Model,
which leads to mixing of the states. In the Higgs sector, we have two complex scalar 
doublets $H_u = (H_u^+, H_u^0)$ and $H_d = (H_d^0, H_d^-)$. The Higgs potential
has  a minimum for non-zero $H_u^0 = v_u$ and $H_d^0 = v_d$, and the ratio between
them is called $\tan \beta$:

\begin{equation}
\tan \beta \equiv v_u/v_d
\end{equation}
When the electroweak symmetry is broken, the Higgs doublets mix with the 
gauge bosons, to give masses to the $W^\pm$ and $Z^0$ bosons. Thus, from
the eight original degrees of freedom only five remain. They form the five scalar
boson states: CP-odd neutral scalar $A^0$, two charged scalars $H^+, H^-$,
and two CP-even scalars $h^0$ and $H^0$.
It can be shown that $h^0$ is the lightest among these Higgs bosons of the MSSM. 

Electroweak symmetry breaking causes the mixing among the sparticles
and produces new mass eigenstates. Neutral higgsinos $\tilde H^0_u, \tilde H^0_d$ 
and neutral gauginos $\tilde B, \tilde W^0$ produce four neutral states called {\em neutralinos} denoted $\chi^0_i$
In a similar way, the charged higgsinos and winos form four charged states,
called {\em charginos} -- two with positive and two with negative electric charges.

In the MSSM we can define a new multiplicative quantum number, called $R-$parity. 
It is defined as:

\begin{equation}
P_R = (-1)^{3(B-L)+2s},
\end{equation}
where $B$ is the barion number, $L$ is the lepton number, and $s$ is the spin of the particle.
The Standard Model particles and the Higgs bosons have even $R-$parity ($P_R=1$), while all
sparticles have odd $R-$parity ($P_R=-1$). The conservation of the $R-$parity has important
phenomenological consequences. It implies that the lightest supersymmetric particle (LSP)
with $P_R=-1$, must be stable, and all sparticles must eventually decay into a state
which contains an odd number of LSPs.

One of the best candidates for the LSP is the lightest neutralino $\chi^0_1$. Since it is electrically neutral and
interacts only weakly with ordinary matter, it becomes a good candidate for dark matter
required by cosmology. Also, in the collider experiments, particles decaying into neutral LSPs
can give an invisible signature (missing energy) in the detector.

\section{Experimental Signals for Supersymmetry}

Although supersymmetry helps solve many theoretical and phenomenological
problems beyond the Standard Model, there is still no experimental evidence 
that supersymmetry really exists. No sparticles have been discovered so far.
The LEP and Tevatron experiments have only set limits to the masses of supersymmetric particles~\cite{Gianotti:vg}.
E.g. the lightest Higgs boson $h^0$ has been excluded in the mass range below 91.0 GeV.

\begin{figure}[ht]
\begin{center}
\includegraphics[width=0.55\textwidth]{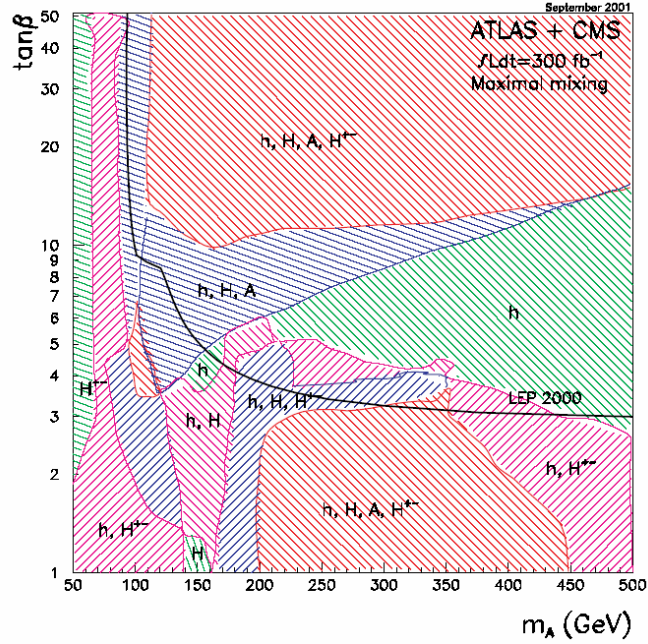}
\caption{LHC discovery potential for the MSSM Higgs bosons \cite{Gianotti:vg}
\label{fig:lhc-mssm-potential}
}
\end{center}
\end{figure}

\begin{figure}[t]
\begin{center}
\includegraphics[width=0.55\textwidth]{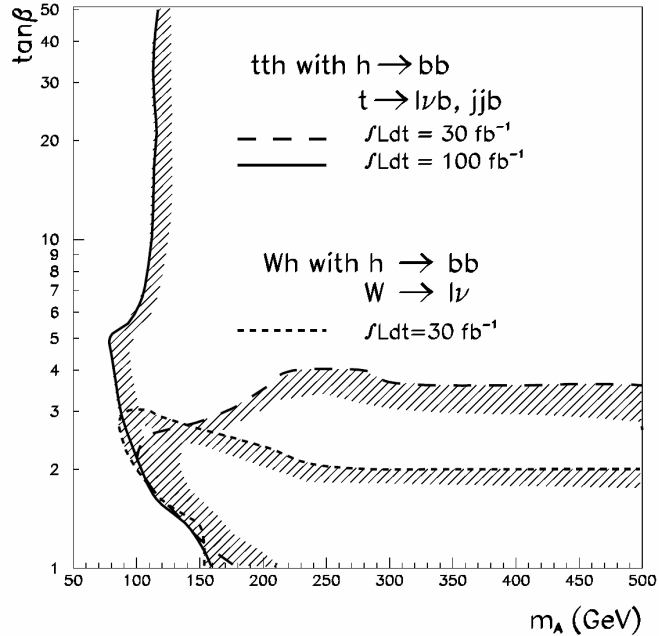}
\caption[The $5\sigma$-discovery contour curves for the $t\bar t$ and $Wh$]{The $5\sigma$-discovery contour curves for the $t\bar t$ and $Wh$ with $h \to b \bar b$ channel
in the $(m_A, tan\beta)$ plane for the ATLAS experiment at luminosities of 30 fb$^{-1}$ \cite{atlas-tdr}.
\label{fig:atlas-tth-bb}
}
\end{center}
\end{figure}

There are plans to search for supersymmetry at the LHC. 
Most of the sparticles with masses below 1TeV may be within reach of this collider.
Fig.~\ref{fig:lhc-mssm-potential} shows the regions in the $(m_A, \tan\beta)$
MSSM parameter space, where the various Higgs bosons can be observed.
Most probable is the detection of the neutral $h$ boson,
which is the lightest of the Higgs particles. Fig.~\ref{fig:atlas-tth-bb}
shows the regions, where $h$ can be observed in the $t\bar t h, h\to b\bar b$ channel.
The possibility to observe the lightest Higgs boson in the invisible decay
is studied in the following chapters of this thesis. 

\chapter{Analysis of $t\bar{t}H$ Production}
\label{chap:analysis}
\section{Introduction}

In this chapter we describe the analysis of the process that 
may allow us to observe the Higgs boson in the invisible channel.
We describe the specific process and study possible backgrounds.
In subsequent sections, we will provide details on the selection criteria 
that were applied to extract the interesting signal events from the background.
The results and possible improvements are presented at the end of this chapter.

The possibilities to observe the Higgs boson in the invisible channel
have been discussed throughout the past years. The prospects of detecting 
such a Higgs particle via its associated production with a gluon, $Z$ or $W$ bosons
are described in~\cite{DPRoy94}.
The exploitation of associated $t\bar{t}$ production was suggested in \cite{Gunion94}.
The weak boson fusion process leading to the production of invisible Higgs is discussed
in \cite{Dieter00}.
A more recent evaluation of the production with gauge bosons can be found in \cite{Godbole03}.

\begin{figure}[ht]
\begin{center}
\includegraphics[width=0.3\textwidth]{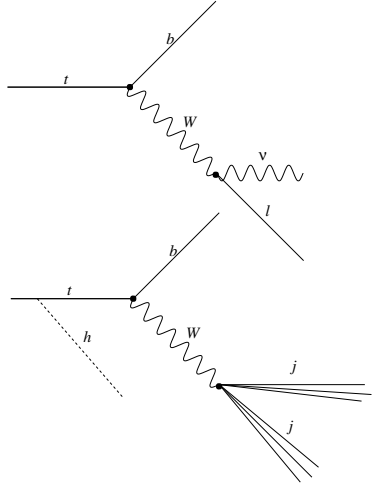}
\label{fig:tthiggs}
\caption{Process for observing the invisible Higgs.}
\end{center}
\end{figure}

In this thesis we study the associated $t\bar{t}H$ production as suggested in \cite{Gunion94}.
The proposed approach was to require the leptonic decay of one of the $W$ bosons
and the hadronic decay of the second one, as shown in Fig.~\ref{fig:tthiggs}.
The final products are a single lepton, 2 $b$-jets and 2 jets from 
the hadronic decay of the $W$ boson. These jets and leptons
can be identified by the detector so they can be used to trigger
the filtering of interesting events.
The evidence of the invisibly-decaying Higgs boson will be an excess of the 
selected events over the predicted background.

\section{Signal and Background Processes}

In order to be able to observe such an excess of signal events, it is necessary 
to simulate and analyze the signal and possible backgrounds. The signal was simulated
using the PYTHIA \cite{pythia} Monte Carlo event generator. The generator simulated 
the proton--proton collisions at the LHC, at the 14 TeV center-of-mass energy.

\noindent$\mathbf{gg, q \bar q \to t \bar t H }$ :

This is the signal process 
generated using the PYTHIA event generator. The simulation is configured
in such a way that 100\% of the Higgs particles are decaying into the 
{\em invisible} channel, meaning they are not observed by the detector.
The simulation assumes the Standard Model coupling of the Higgs to the $t$ quarks.

The background may come from processes which give 
a signature that is similar to the one generated by the signal process, i.e. an isolated lepton,
2 identified $b$-jets, 2 additional jets and missing transverse energy.
The processes that may lead to such signatures, are the $t$-quark or $b$-quark
production associated with the $W$- or $Z$-boson.

\noindent$\mathbf{gg, q \bar q \to t \bar t}$: 

This is the main 
background process. It can yield the same signature as the signal,
but the main difference is in the missing energy. In the case of
the signal, it contains the energy of the invisible Higgs, 
while in the background it mostly carries the energy of the
neutrino from the leptonic decay of the $W$ boson.
Consequently, the missing energy will be the used to reduce this background.

\noindent$\mathbf{gg, q \bar q \to t \bar t Z, Z \to \nu \nu}$ :

This background can yield the same signature as the signal. The neutrinos 
coming from the $Z$ decay can carry significant missing energy, comparable with the signal.

\noindent$\mathbf{q \bar q \to t \bar t W, W \to \ell \nu}$ :

In this process, there can be 3 $W$ bosons: one from the associated production of the $t \bar t$
pair and 2 from the top-quarks decays. As a consequence, the results
of the simulation where only the $W$-boson from the associated production
 is forced to decay into a lepton and the neutrino,
should be finally multiplied by a combinatorial factor of 3. It assumes that the 
acceptance of the analysis is independent of the source of the lepton.

\noindent$\mathbf{gg, q \bar q \to  b \bar b Z/\gamma^*, Z/\gamma^* \to \ell \ell \oplus {\mathbf \rm  jets}}$ :

This process can yield the same signature as the signal and its cross-section is several orders of
magnitude larger. However, it is possible to reduce this background considerably 
by requiring the reconstruction of the top-quark from the hadronic jets.
Also, requiring significant missing energy and vetoing additional leptons 
will suppress this background.

\noindent$\mathbf{q \bar q \to b \bar b W, W \to \ell \nu \oplus {\mathbf \rm
jets}}$ :

This background can be suppressed by requiring the reconstruction
of the top quark in the hadronic mode and significant transverse energy.

\section{Event Generation and Detector Simulation}

For the $t \bar t$ process event generation, the 
PYTHIA \cite{pythia} and HERWIG \cite{Herwig} Monte Carlo generators were used.
Additionally, the AcerMC \cite{AcerMC} event generator was used to simulate
other backgrounds. For these generators the initial parameters were
set to default values.

For detector simulation, the ATLFAST \cite{ATL-PHYS-98-131} program was used. 
It implements a simplified simulation of the ATLAS detector 
at the LHC. The generated events serve as input to this program 
and it produces the detector response as output. 
It provides such observables as the 4-momenta of isolated leptons,
jets, tagged $b$-jets and transverse energy. It is also possible
to evaluate the missing transverse energy by integrating the 
total energy of particles detected and subtracting from the initial
energy of colliding protons. 

The simulation assumes 90\% efficiency in lepton
identification, 80\% for jet reconstruction and 60\%
for $b$-jet identification (tagging). These efficiencies
were taken into account while calculating
the normalized number of events. 

\section{Analysis}

The objective of the analysis is to apply such filters
to the simulated processes that will suppress,
as much as possible, the background while preserving a high 
number of signal events.
The following filters are applied to the events:

\begin{figure}[ht]
\begin{center}
\includegraphics[width=0.45\textwidth]{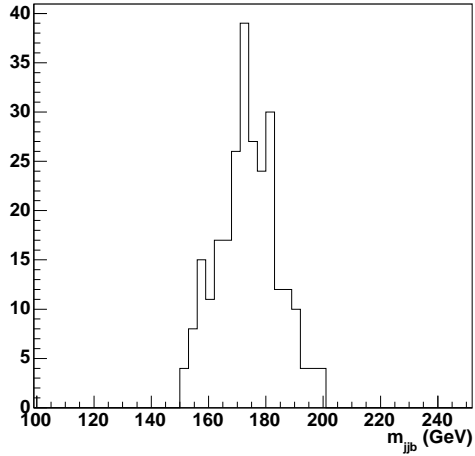}
\caption[Reconstructed mass of the top quark for the ttH events]{Reconstructed mass of the top quark for the ttH events (number of events not normalized).
\label{fig:mtrecbest}
}
\end{center}
\end{figure}

\begin{figure}[htb]
\begin{center}
\includegraphics[width=0.45\textwidth]{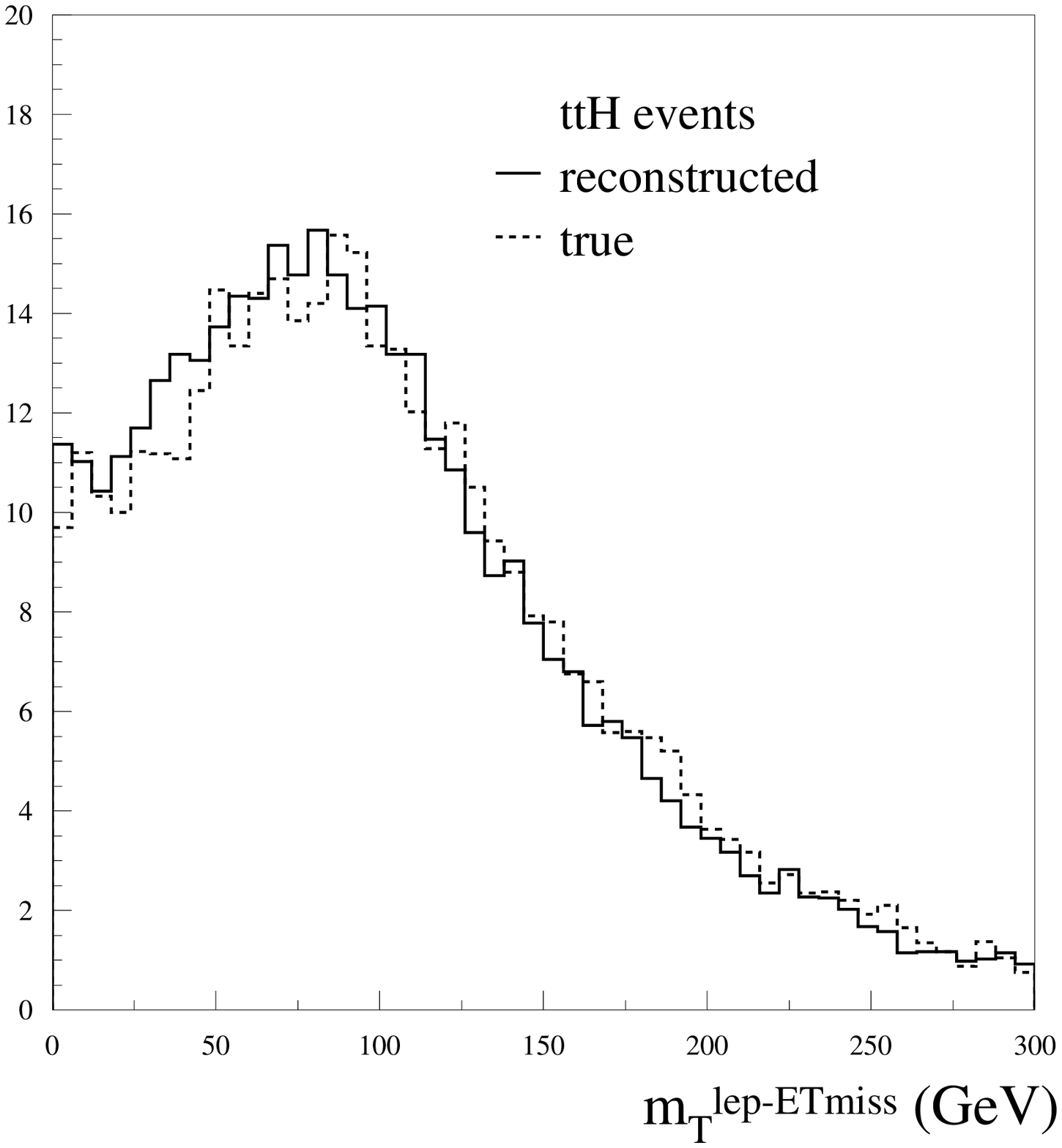}
\includegraphics[width=0.45\textwidth]{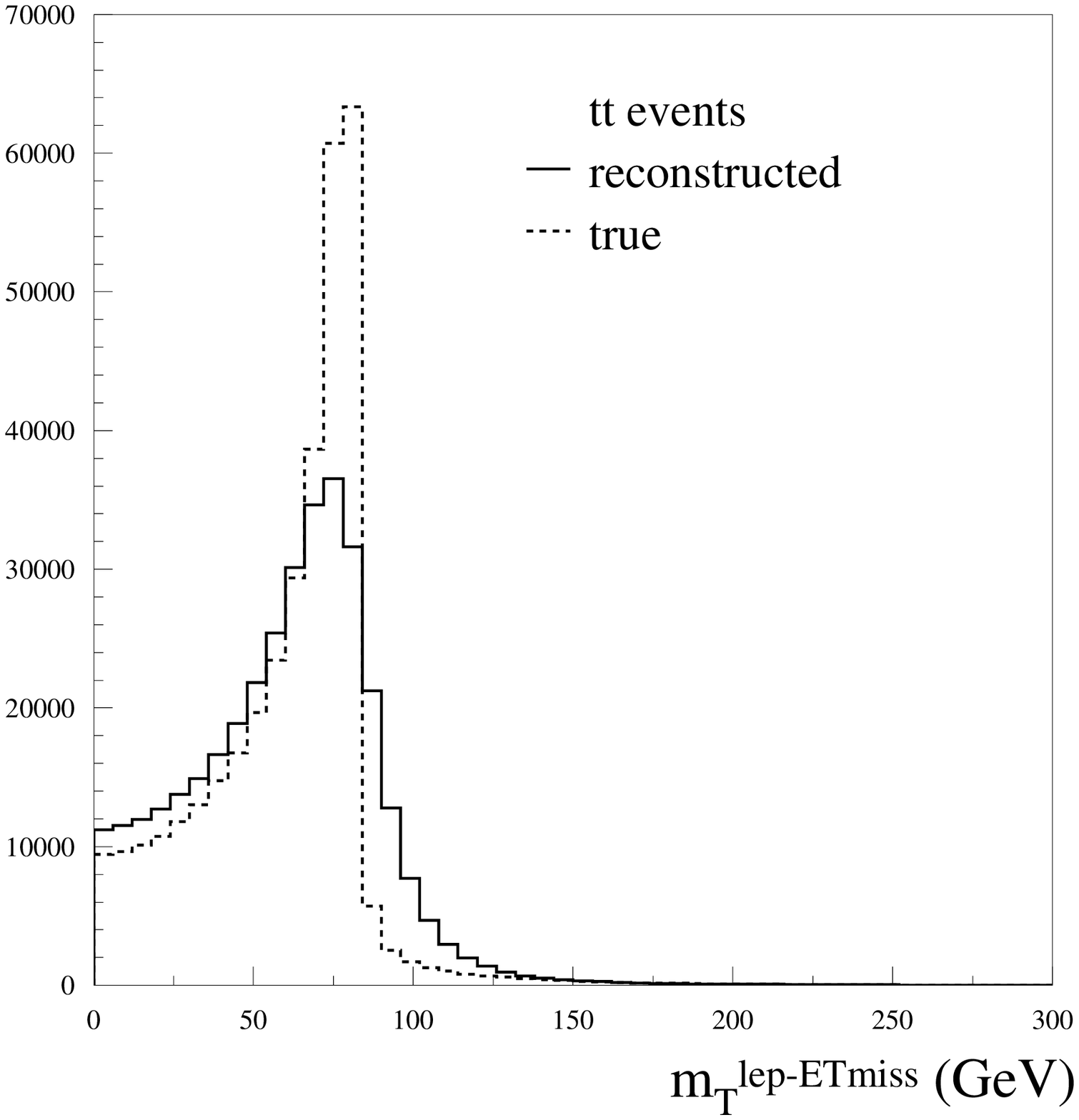}
\caption
[Reconstructed transverse mass of the lepton and \etmiss~ system]
{
\small
\label{fig:mtrans}
Reconstructed transverse mass of the lepton and \etmiss~ system in the $t
\bar tH$ events (left plot) and in the $t \bar t$ events (right plot).  The
dashed line denotes the distributions calculated from the true invisible energy
of the primary products of W boson decays in these events, obtained by using the
generator level information.  The distributions are normalized to the number of
events expected for an integrated luminosity of $30 fb^{-1}$.}
\end{center}
\end{figure}

\begin{enumerate}
\item
One isolated lepton is required. The lepton comes from semi-leptonic
decay of one top quark. Events with more leptons
are rejected in order to eliminate events coming from the $b \bar b Z / \gamma^*$
background.

\item
Two identified (tagged) $b$-jets are required.

\item
One top quark is required to be reconstructed from the hadronic decay mode
$ t \to jjb $. First, only such $jj$ pairs are selected from all possible
combinations of jets, where $m_{jj} = m_W \pm$~15~GeV. Additionally,
only central jets with pseudo-rapidity $|\eta^{jet}| < 2.0$ are
allowed to reduce the number of events with jets coming from
initial- or final-state QCD radiation instead of the $W$-boson decay.
Next, the masses of all combinations of the selected $jj$ pairs and $b$-jets 
are calculated. The four-momenta of the jets are recalibrated 
so that the $m_{jj}$ gives $m_W$ exactly. The best combination
of the $jjb$ system is chosen and the top quark is considered reconstructed if
$m_{jjb}=m_t \pm 25$~GeV (see Fig.~\ref{fig:mtrecbest}).  

\item
It is not possible to reconstruct the second top-quark which decays in
the leptonic channel. This would require precise information on the missing energy,
which in the case of the $t \bar t H $ events comes both from the leptonic decay of the top-quark
and from the invisible Higgs. Instead, it is possible to use the transverse mass of the 
lepton end \etmiss~ system to distinguish between the signal and the $t \bar t$ background, where 

\begin{equation}
m_{T}(\ell,\etmiss)=\sqrt{2E_T^\ell \etmiss(1-cos\phi)}
\end{equation}
The Fig. \ref{fig:mtrans} shows the distribution of $m_{T}(\ell,\etmiss)$ for signal and background processes.
It can be seen that for the $t \bar t$ process, where the missing energy comes from
the leptonic decay of the $W$-boson, there is a sharp end of the distribution 
around the $W$ mass. To cut the big fraction of the $t \bar t$ 
events we require that $m_{T}(\ell,\etmiss) > 120$~GeV.

\item 
In addition to the $m_{T}(\ell,\etmiss)$ cut, we also require a large missing transverse
energy of the event: $\etmiss > 150$~GeV.

\item

The signal-to-background ratio is enhanced by the additional
requirement of large transverse momenta in the reconstructed system, $ \sum
p_T^{\mathrm{rec}} > 250$~GeV. The $ \sum p_T^{\mathrm{rec}}= \sum p_T^j + p_T^l$ where the sum
runs over the transverse momenta of reconstructed objects from top-quark
decays: two $b$-jets, two light jets used for the reconstruction of the $W \to q
\bar q$ decay and an isolated lepton. This further suppresses the background
where true top quarks are not present, like $b \bar b Z$ and $b \bar b W$.
\item
Finally, further enhancement of the signal-to-background ratio
can be achieved by the additional
requirement regarding cone separation, $R_{jj}$, between jets used for the
$W \to jj$ reconstruction, the $R_{jj} < 2.2$.

\begin{equation}
R_{jj} = \sqrt{(\Delta \eta)^2 + (\Delta \phi)^2}
\end{equation}
is a distance in the $(\eta, \phi)$ plane between the jets.

\end{enumerate}

Table \ref{T3:a} shows the cumulative acceptance of the specified filters
applied to selected processes. 

\begin{table}[htb]
\small
\newcommand{\lstrut}{{$\strut\atop\strut$}}
  \caption[The cumulative acceptances for the specified selection
  criteria]{\small The cumulative acceptances for the specified selection
  criteria. Efficiencies for b-tagging and lepton identification are
  included. A Higgs boson mass of 120 GeV is assumed for signal events. 
  Only the dominant background sources are listed.
}
\label{T3:a}
\vspace{2mm}
\begin{center}
\begin{tabular}{lllll}
\hline\noalign{\smallskip}
Process & $t \bar t H$ & $t \bar t Z$ & $t \bar t$  & $t \bar t$  \\
        &\hspace{-0.3cm} {\tt PYTHIA} &\hspace{-0.3cm} {\tt AcerMC}&\hspace{-0.3cm}  {\tt PYTHIA} &\hspace{-0.3cm} {\tt HERWIG} \\
\noalign{\smallskip}\hline\noalign{\smallskip}
    Trigger lepton &\hspace{-0.3cm} 22\%  &\hspace{-0.3cm} 22\%  &\hspace{-0.3cm} 22\%  &\hspace{-0.3cm} 22\%     \\
\noalign{\smallskip}\hline\noalign{\smallskip}
2 b-jets + 2 jets &\hspace{-0.3cm} 5.0\%  &\hspace{-0.3cm} 4.8\% &\hspace{-0.3cm} 4.9\% &\hspace{-0.3cm} 5.2\%    \\
\noalign{\smallskip}\hline\noalign{\smallskip}
rec. t-quark (jjb) &\hspace{-0.3cm}  2.6\%  &\hspace{-0.3cm} 2.4\% &\hspace{-0.3cm} 2.4\% &\hspace{-0.3cm} 2.6\%     \\
\noalign{\smallskip}\hline\noalign{\smallskip}
$m_T^{\ell,\etmiss}>120$ GeV &\hspace{-0.3cm} 0.87\%  &\hspace{-0.3cm} 0.93\% &\hspace{-0.3cm} $4.1 \cdot 10^{-4}$ &\hspace{-0.3cm}  $5.2 \cdot 10^{-4}$  \\
\noalign{\smallskip}\hline\noalign{\smallskip}
$\etmiss > 150$ GeV &\hspace{-0.3cm} 0.41\%  &\hspace{-0.3cm} 0.53\% &\hspace{-0.3cm} $ 2.3 \cdot 10^{-5}$ &\hspace{-0.3cm} $ 3.7 \cdot 10^{-5}$  \\
\noalign{\smallskip}\hline\noalign{\smallskip}
$\sum p_T^{\mathrm{rec}}>250$ GeV &\hspace{-0.3cm} 0.40\%  &\hspace{-0.3cm} 0.51\% &\hspace{-0.3cm}  $ 2.0 \cdot 10^{-5}$ &\hspace{-0.3cm} $ 3.2 \cdot 10^{-5}$ \\
\noalign{\smallskip}\hline\noalign{\smallskip}
$R_{\mathrm{jj}}<2.2$ &\hspace{-0.3cm} 0.28\%  &\hspace{-0.3cm} 0.35\% &\hspace{-0.3cm}  $ 7.5 \cdot 10^{-6}$ &\hspace{-0.3cm} $ 1.2 \cdot 10^{-5}$  \\
\noalign{\smallskip}\hline
\end{tabular}
\end{center}
\end{table}

We can see that the acceptance of signal events is about 0.3\%,
while the background is reduced by a factor of $10^-5 - 10^-6$.

After performing such selections, one can see that about 
70\% of the $t \bar t$ events come from the lepton-tau
decay and 20\% from the lepton-lepton decay of the top-quark pair in the {\tt PYTHIA}
sample with compatible fractions also found in {\tt HERWIG} events.
The {\it lepton-tau} label denotes one top quark decaying $t
\to Wb \to \ell \nu b$ and another $t \to Wb \to \tau \nu b$, where
$\ell$ stands for electron or muon.
The {\it lepton-lepton} decay labels events
 with both top quarks decaying  $t\to Wb \to \ell \nu b$. Finally,
 the {\it lepton-hadron} decay labels events with one top quark decaying $t
\to Wb \to \ell \nu b$ and another $t \to Wb \to q \bar q  b$.

\begin{figure}[htb]
\begin{center}
\includegraphics[width=0.45\textwidth]{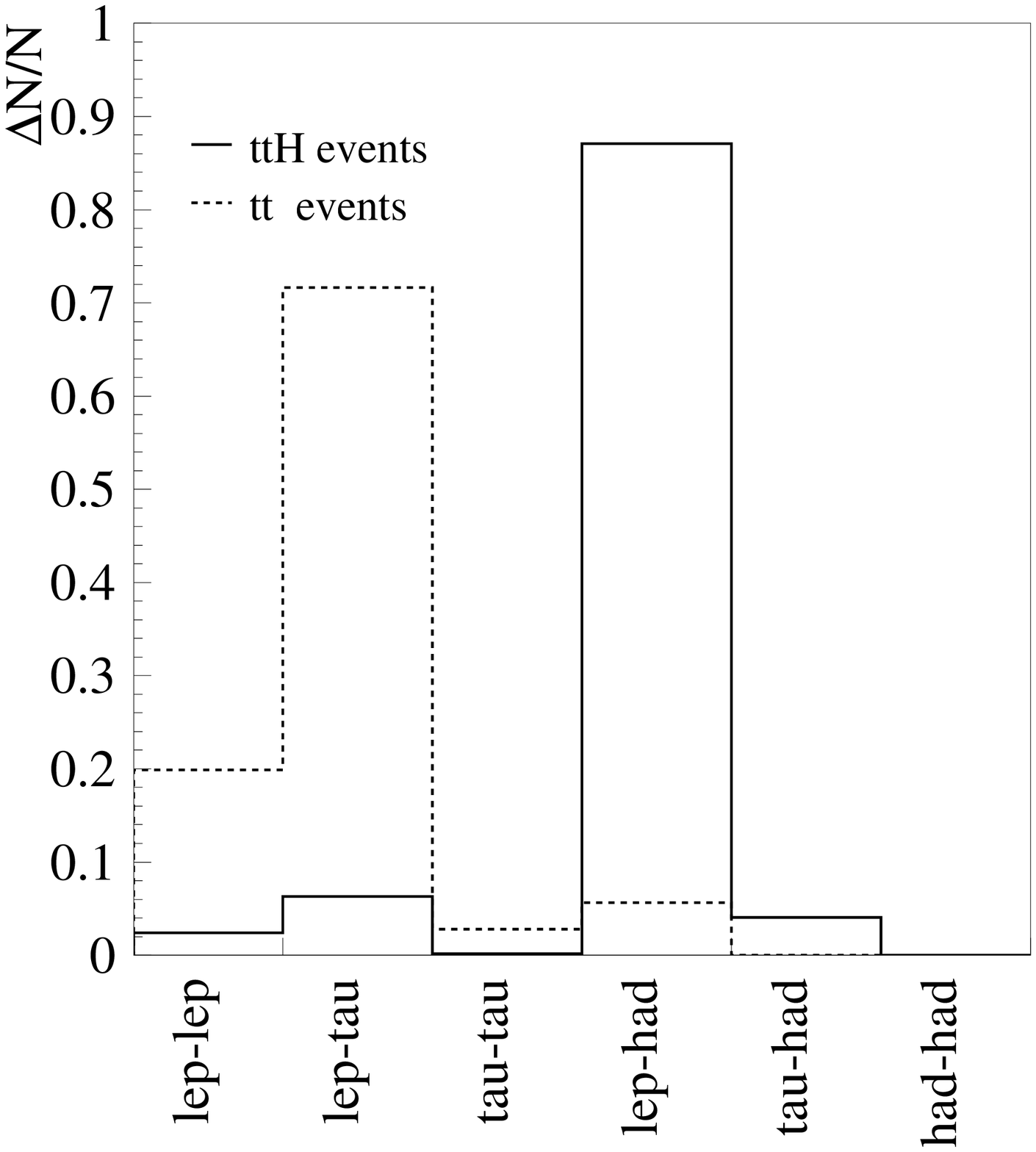}
\includegraphics[width=0.45\textwidth]{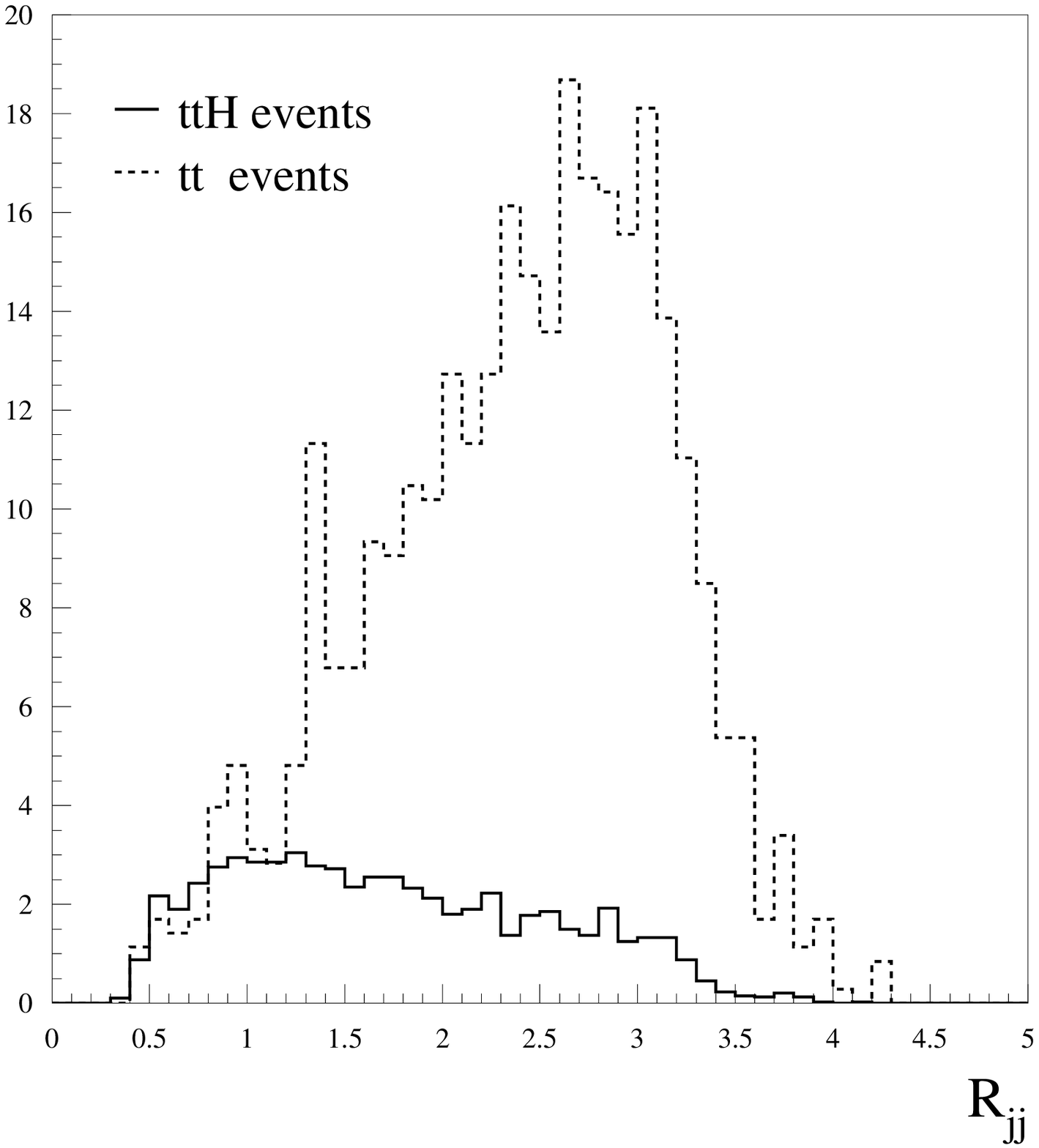}
\label{fig:rec}
\caption[The decay modes for the signal and background and the cone separation]
{
\small
The relative fractions of the $t \bar t$ decay modes are listed for signal and $t \bar t$
background simulated with {\tt PYTHIA} (left plot).
The $R_{\mathrm{jj}}$ cone separation between jets used in the $W \to jj$ reconstruction;
the distributions are normalized to the number of events expected for
 an integrated luminosity of $30 fb^{-1}$ (right plot).
}
\end{center}
\end{figure}

In the cases of lepton-lepton and lepton-tau decays, the $jjb$ combination 
is thus made not from the true $W \to q \bar q$ decays,
but from other jets coming from the initial- or final-state radiation (ISR/FSR).
  These events could
hopefully be suppressed further by implementing a tau-jet veto and with more
stringent requirements in the $t \to jjb$ reconstruction. 
The signal events, in
contrast, contain only a $\sim 10\%$ fraction of lepton-tau and lepton-lepton decays. The
relative fractions of signal and $t \bar t$ background events are shown in Figure~\ref{fig:rec}.

Considering the relative fractions of the tau-lepton events in the signal and background,
one can assume that the inter-jet cone separation, $R_{\mathrm{jj}}$, might provide
some additional separation power; the $R_{\mathrm{jj}}$ for signal and $t \bar t$ background
events are given in Figure~\ref{fig:rec}. Subsequently, a loose cut of $R_{\mathrm{jj}}<2.2$ was
applied; the final efficiencies are listed in Table~\ref{T3:a}.

\begin{table}[htb]
\small
  \caption[Expected number of events after applying the filters] 
  {\small Expected numbers of events for an integrated luminosity
  of $30 fb^{-1}$ and selection as specified in Table~\ref{T3:a}.
  Efficiencies for b-tagging and lepton identification
  are included. The {\tt (PY)} and {\tt (HW)} denote the results for the
  $t \bar t$ events generated with {\tt PYTHIA} and {\tt HERWIG}
  respectively. Also shown is the separate contribution to the $t \bar
  t$ background from the lepton-hadron events.
\label{T3:b}}
\vspace{2mm}
\begin{center}
\begin{tabular}{lc}
\hline\noalign{\smallskip}
Process & No. of events  \\
\noalign{\smallskip}\hline\noalign{\smallskip}
$t \bar t H$,               &         \\
              $m_H=100$ GeV &  60     \\ 
              $m_H=120$ GeV &  45     \\ 
              $m_H=140$ GeV &  30     \\ 
              $m_H=160$ GeV &  25     \\ 
              $m_H=200$ GeV &  15     \\ 
\noalign{\smallskip}\hline\noalign{\smallskip}
$t \bar t Z$                 &  20      \\ 
\noalign{\smallskip}\hline\noalign{\smallskip}
$t \bar t W  $              &   20  \\ 
\noalign{\smallskip}\hline\noalign{\smallskip}
$t \bar t $ (all)                 &  115 (PY) ,  190 (HW)    \\ 
 (only lepton-hadron)                &   15 (PY) ,   30 (HW)  \\ 
\noalign{\smallskip}\hline\noalign{\smallskip}
$b \bar b W $               &   5 \\ 
\noalign{\smallskip}\hline\noalign{\smallskip}
$b \bar b Z/\gamma^*$       &   5  \\ 
\noalign{\smallskip}\hline
\end{tabular}
\end{center}
\end{table}

The expected numbers of events for an integrated luminosity of $30 fb^{-1}$ are
given in Table~\ref{T3:b}. Several values of the Higgs boson masses are studied,
while assuming the Standard Model production cross-sections.
The branching ratio of the Higgs boson to the invisible channel is assumed to be 100\%
in this analysis.

It can be seen that for the process $t \bar t$ generated with PYTHIA,
the signal-to-background ratio is about 39\% for
a Higgs boson mass of 100 GeV and about 9\% for a Higgs mass of 200~GeV.
Of course, if only true reconstructed top-quark decays were allowed,
these ratios would increase considerably.

It would also be possible to increase the signal-to-background ratio
by setting more tight cuts on such observables as
$R_{\mathrm{jj}}$, \etmiss, $\sum p_T^{\mathrm{rec}}$. 
However, such optimizations are not suitable at this moment,
since they could lead to over-fitting to the results obtained by the 
particular implementations of the processes in event generators.
It is clear that further optimization of the analysis should 
lead to an increase in the efficiency of rejecting "fake" $W \to jj$
reconstructions and eliminating the contribution of 
events with lepton-tau and lepton-lepton decays.

The basic measure of the quality of the analysis is the significance $S=s/\sqrt{b}$,
where $s$ and $b$ are the numbers of signal and background events respectively. $S=5$ is assumed
to be the {\em discovery} level, while $S=3$ is called the {\em evidence} level.
$S=1.96$ corresponds to the 95\% confidence level.
We can define the $\xi^2$ parameter as:
$$\xi^2 = \frac{\sigma{\rm (t \bar t H)}}
{\sigma{\rm (t \bar t H)}_{\rm SM}} \times {\rm Br(H \to inv)}.$$
It can be used to estimate the minimal branching ratio required 
for the significance to reach the desired thresholds. 

Taking the values from Table~\ref{T3:b} we get $s = 60$ for the $t\bar{t}H$ signal
with $m_{H}=100 GeV$ and $b = 65$ for the total background that includes 
the (lep-had) $t\bar{t}$ events generated with PYTHIA.
This yields $S=7.44$ and requires $\xi^2$ to be
{\bf 0.67} and {\bf 0.4} for the discovery and evidence levels respectively.
The values for the $t\bar t$ background and also for (lep-had) events
are shown in Table~\ref{tab-sigma}.

\begin{table}[htb]
\begin{center}
\caption[Estimated significance and branching ratio]{Estimated significance $S = s/\sqrt{b}$ for $t \bar t$ background and estimated branching ratios required
for significance to reach the values of 5, 3 and 1.96
\label{tab-sigma}
}
\vspace{2mm}
\begin{tabular}{l|cccc|cccc}
\hline
  & all $t \bar t $ &  & &  & lep-had $t \bar t $ &  & & \\
    $m_H$          & $s/\sqrt{b}$, & $\xi^2 (5)$ & $\xi^2(3)$ & $\xi^2(1.96)$ & $s/\sqrt{b}$ &$\xi^2(5)$  &$\xi^2(3)$  &$\xi^2(1.96)$ \\

\hline
100 GeV	&4.67	&1.07&	0.64&	0.42&	7.44&	0.67&	0.4&	0.26\\
120 GeV	&3.5	&1.43&	0.86&	0.56&	5.58&	0.9&	0.54&	0.35\\
140 GeV	&2.34	&2.14&	1.28&	0.84&	3.72&	1.34&	0.81&	0.53\\
160 GeV	&1.95	&2.57&	1.54&	1.01&	3.1&	1.61&	0.97&	0.63\\
200 GeV	&1.17	&4.28&	2.57&	1.68&	1.86&	2.69&	1.61&	1.05\\
\hline
\end{tabular}
\end{center}
\end{table}

\chapter{Scans of mSUGRA Model}
\label{chap:scans}
\section{Objective of the scans}

The supersymmetric models as described in Chapter~\ref{chap:susy} allow a decay of the Higgs boson
into a pair of the lightest neutralinos $\chi^0_1 \chi^0_1$. This is an invisible channel
when the neutralino is the LSP, which means it does not decay, but leaves the detector,
leaving its signature only as missing energy. 
It is interesting to study for which parameters of the supersymmetric models, like mSUGRA,
such decay is allowed and can lead to possible Higgs discovery.
Results of such scans can be combined with results from similar scans for Higgs
decaying into visible Standard Model particles.

In this thesis, the {\tt Suspect} and {\tt HDECAY} \cite{suspect,Djouadi:1997yw} 
programs were used to perform the scan 
of the mSUGRA model parameter space. The goal of the scan was to identify the regions with
high Higgs branching ratio to $\chi^0_1\chi^0_1$ pair. 
The analysis of the associated $t \bar t H$ production presented in \cite{epj03}
and discussed in Chapter~\ref{chap:analysis} 
was performed with the assumption that the $BR(H \rightarrow \chi^0_1\chi^0_1)$ was 100\%.
Therefore, it is necessary to multiply the results of the analysis by the 
branching ratio obtained from the scans. 
The estimates for minimal branching ratios 
required for the discovery or evidence signatures respectively
are included in Table~\ref{tab-sigma}.
In the scans, we will search for such areas in the parameter space, where the
$BR(H \to \chi^0_1 \chi^0_1)$ can reach these values.  

It is also interesting to check the branching ratio of Higgs
to the $b \bar b$ pair. This decay is considered as the main channel
where the Higgs can be observed in the same production mode. 
If the decay of $H\to\chi^0_1\chi^0_1$ is open, the $H\to b b$ will
become suppressed.
We can compare the branching 
ratio to the $b \bar b$ and the invisible channels and examine if the
invisible decay should be considered as complementary for the observation.

\section{Implementation of the Scans}

For the scans, we have combined two programs. {\tt Suspect}~\cite{suspect} 
takes as input the 5 parameters of the mSUGRA model (see eq. (\ref{eq-msugra}))and returns
the spectrum of masses of MSSM particles. These parameters serve as
input to the {\tt HDECAY}~\cite{Djouadi:1997yw} program, which computes the branching ratios 
of the Higgs to both Standard Model and MSSM particles. 

The input parameters were selected to cover the most interesting
regions of the mSUGRA space. The masses $m_0$ and $m_{1/2}$
were covered in the range of 0 to 500 GeV, with steps of 10 GeV.
The $\tan \beta$ had values of 50, 25 and 12.5, \ldots Three values of the $A_0$
parameter were chosen: -200, 0, 200 and both values of sign~$\mu$ were used.

As the input for the {\tt Suspect} and {\tt HDECAY} programs,
default parameters were used. 
For the drawing of plots, the ROOT framework was used~\cite{root-framework}.

\section{Constraints on the Parameter Space}

It is important to note that not all regions of the mSUGRA parameter space
are allowed. First, it is possible that the mass
of the stau is lower than that of the lightest neutralino. This would imply that
the neutralino is not the LSP, contradicting the assumption
that SUSY particles decay into the invisible channel. Such regions should be theoretically excluded.

Second, current experimental results from the LEP and Tevatron experiments, as well
as from cosmological observations have excluded the existence of SUSY
particles below certain mass limits. Such limits for the SUSY Higgs
boson obtained by four combined LEP experiments is now set at 91.0 GeV \cite{LEP-final}.
More limits on the masses of other sparticles can be found in \cite{Gianotti:vg}.
The constraints applied to the scans are shown in Table~\ref{T3:limits}.

Additionally, the programs that were used for the scans have limitations
related to numerical stability.
Although several tests are performed internally by these programs, there are also points
of the parameter space, where the calculations do not converge. 
Such points are also excluded from the scans.

\begin{table}[htb]
\small
  \caption[Experimental constraints on masses of sparticles used in the scan]
  {\small Experimental constraints on masses of sparticles used in the scan
\label{T3:limits}}
\vspace{2mm}
\begin{center}
\begin{tabular}{ll}
\hline\noalign{\smallskip}
Particle & Mass constraint  \\
\noalign{\smallskip}\hline\noalign{\smallskip}
$h$          &  > 91.0 GeV\\
sleptons     &  > 80.0 GeV\\
chargino     &  > 103.6 GeV\\
gluino       &  > 195.0 GeV\\
neutralino   &  > 45.6 GeV\\
\noalign{\smallskip}\hline
\end{tabular}
\end{center}
\end{table}

\section{Results of the Scans}

\begin{figure}[htb]
  \begin{center}
    \includegraphics[width=0.95\textwidth]{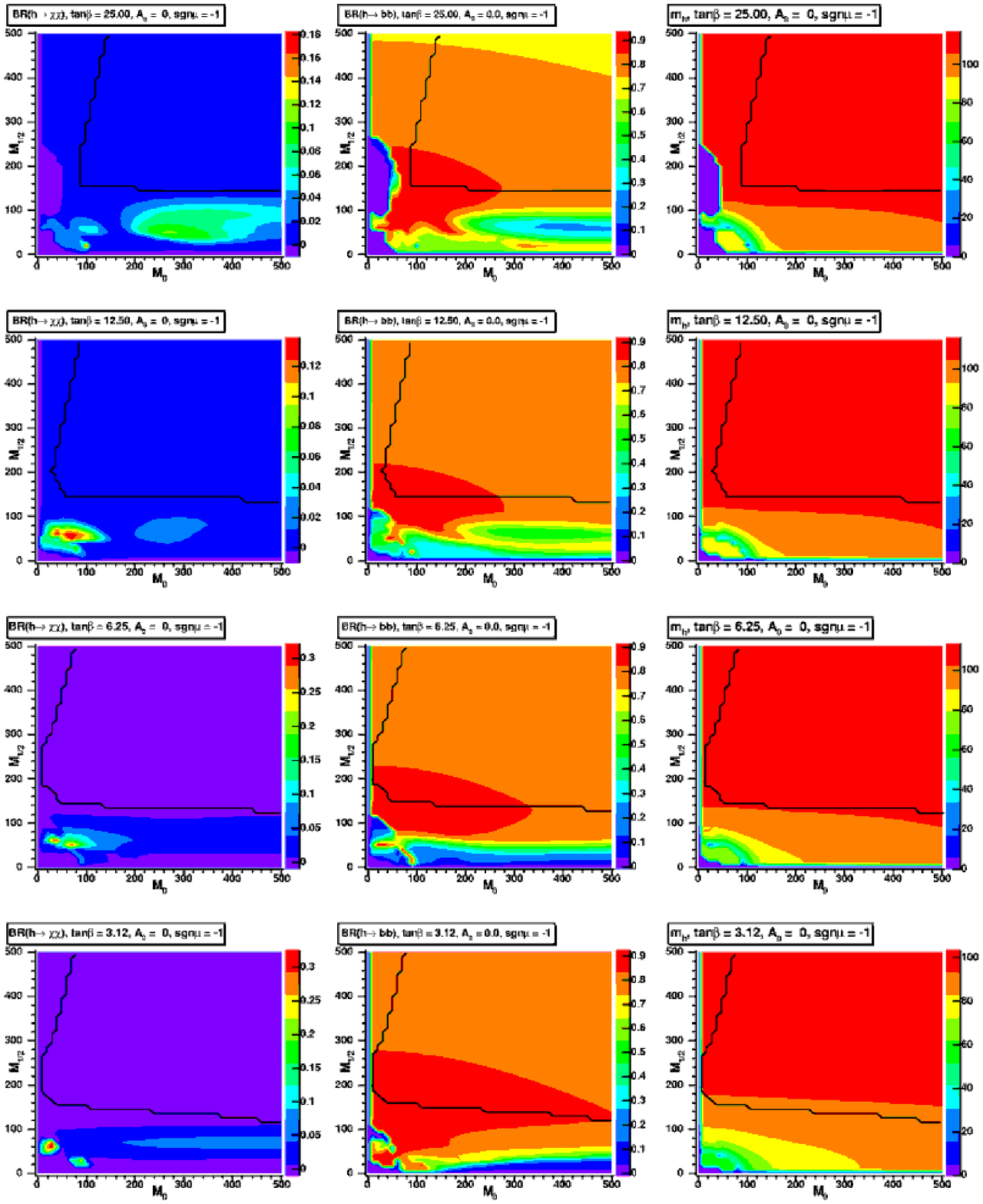}
  \end{center}
  \caption{Scans of mSUGRA parameter space for $\mu < 0$.}
 \label{canvas1}
\end{figure}

\begin{figure}[htb]
  \begin{center}
    \includegraphics[width=0.95\textwidth]{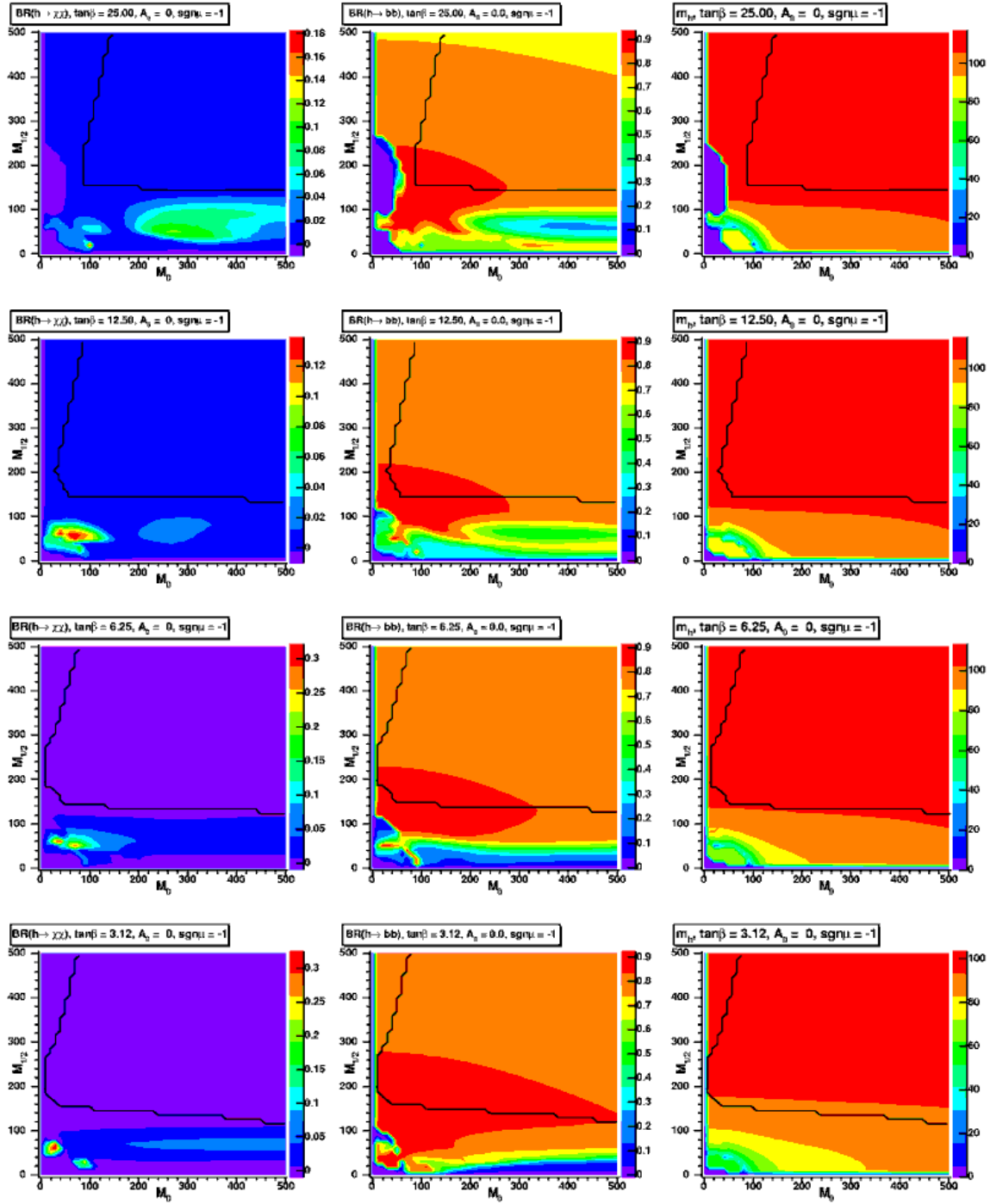}
  \end{center}
  \caption{Scans of mSUGRA parameter space for $\mu > 0$.}
\label{canvas2}
\end{figure}

The resulting plots showing the contours of branching ratios are presented 
in Figs.~\ref{canvas1}-\ref{canvas2}.
We selected the following values of tan$\beta$: 25, 12.5, 6.25, 3.12; both
signs of $\mu$ and produced plots in the $(m_0, m_{1/2})$ plane. We show only values of $A_0=0$,
since other tested values (-200, 200) provide similar results. For $\tan\beta>25$ and $<3$ 
we observed more instability in the combined programs.

For each value of $\tan\beta$ and sign $\mu$ we show three contour plots in the $(m_0, m_{1/2})$ plane:
branching ratio of lightest Higgs boson to neutralinos, to the $b \bar b$ pair and the Higgs mass $m_h$.  
The theoretical and experimental constraints are indicated by the solid line: regions
below and to the left of this line are excluded. 

We can see that there are small regions where the $BR(h \to \chi^0_1 \chi^0_1)$
is dominant. For the $\mu >0$ and for small $\tan\beta$, the value of the BR can reach 0.9
for small regions of the parameter space. This would mean that assuming the 
possibility of analysis as described in Chapter~\ref{chap:analysis} a sufficient excess of signal events would be observed to enable observation of the Higgs
in the invisible channel.
However, taking into account the experimental limits, it is very likely that 
the regions where the $BR(h \to \chi^0_1 \chi^0_1)$ is large, are excluded by current observations. 
On the other hand, the plots show that the branching ratio to $b$-quark 
is dominant in most of the regions that are not excluded. That confirms the need for the 
search for Higgs in this decay channel.

\begin{figure}[htb]
  \begin{center}
    \includegraphics[width=0.43\textwidth]{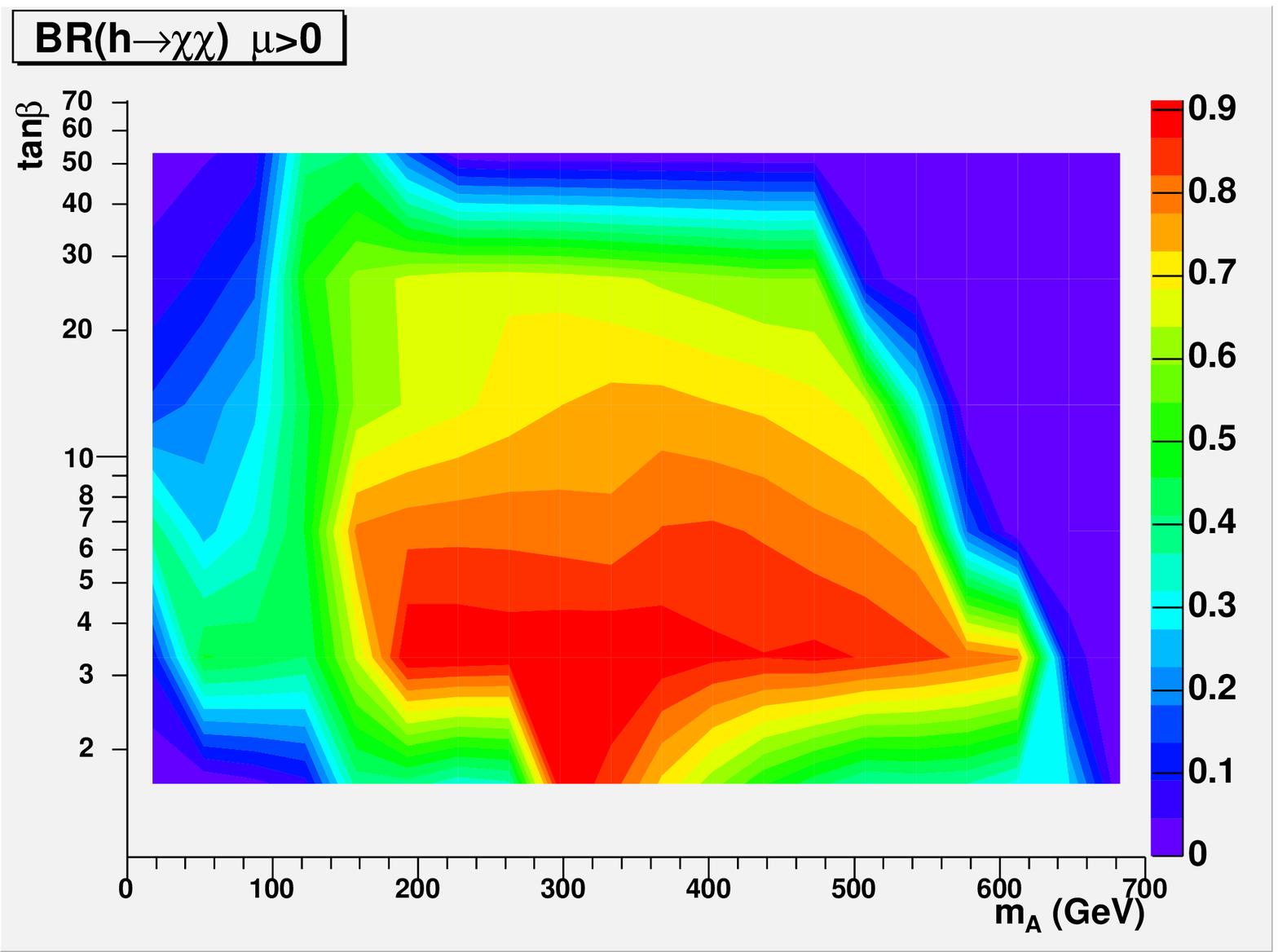}
    \includegraphics[width=0.43\textwidth]{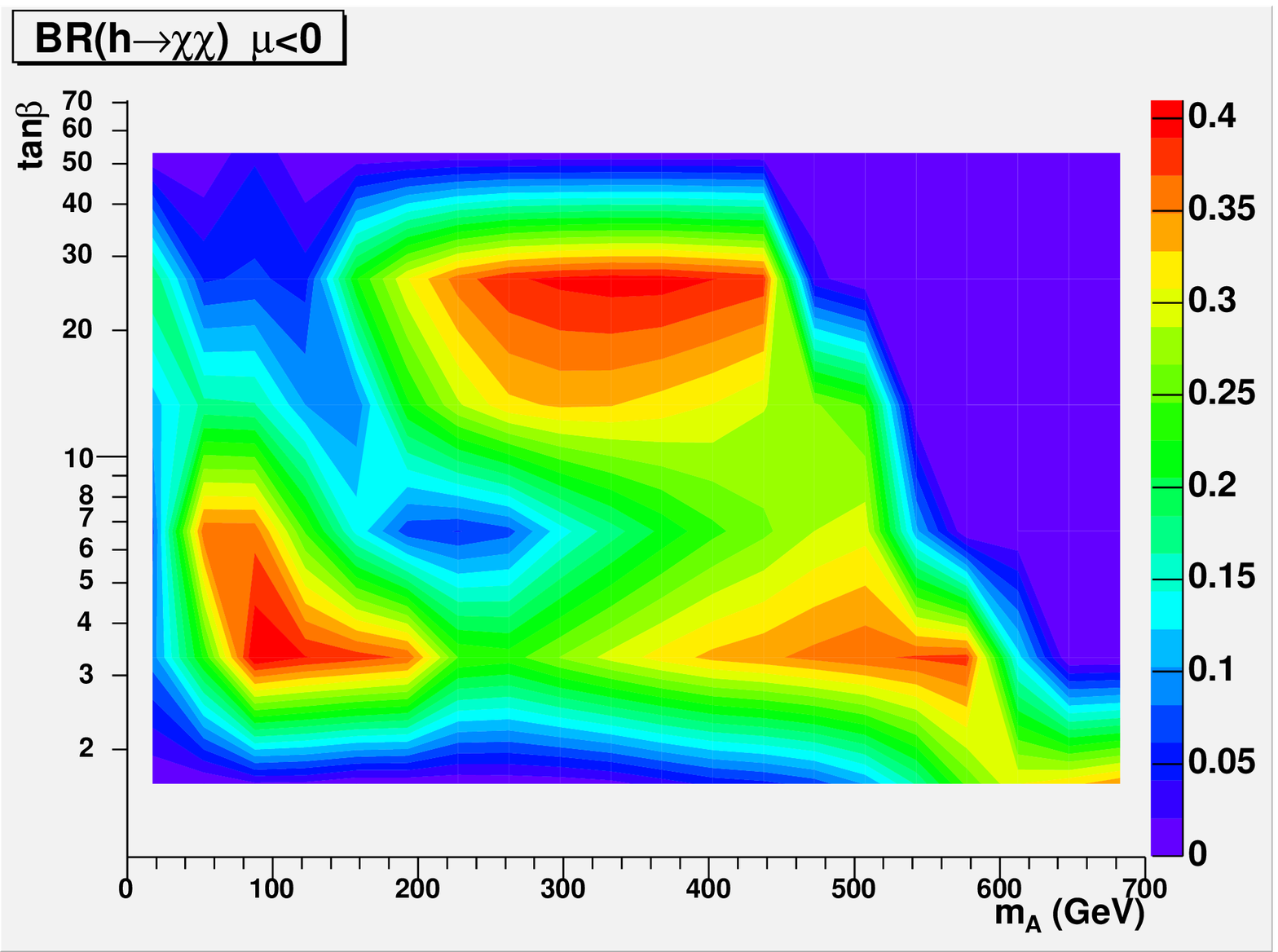}
  \end{center}
  \caption[Scans of mSUGRA model in the $(m_A, tan \beta)$ plane.]{\small $BR(h\to\chi^0_1\chi^0_1)$ 
in mSUGRA model in the $(m_A, tan \beta)$ plane, $A_0=0$, without applying any constraints.}
 \label{figmachi}
\end{figure}

\begin{figure}[htb]
  \begin{center}
    \includegraphics[width=0.43\textwidth]{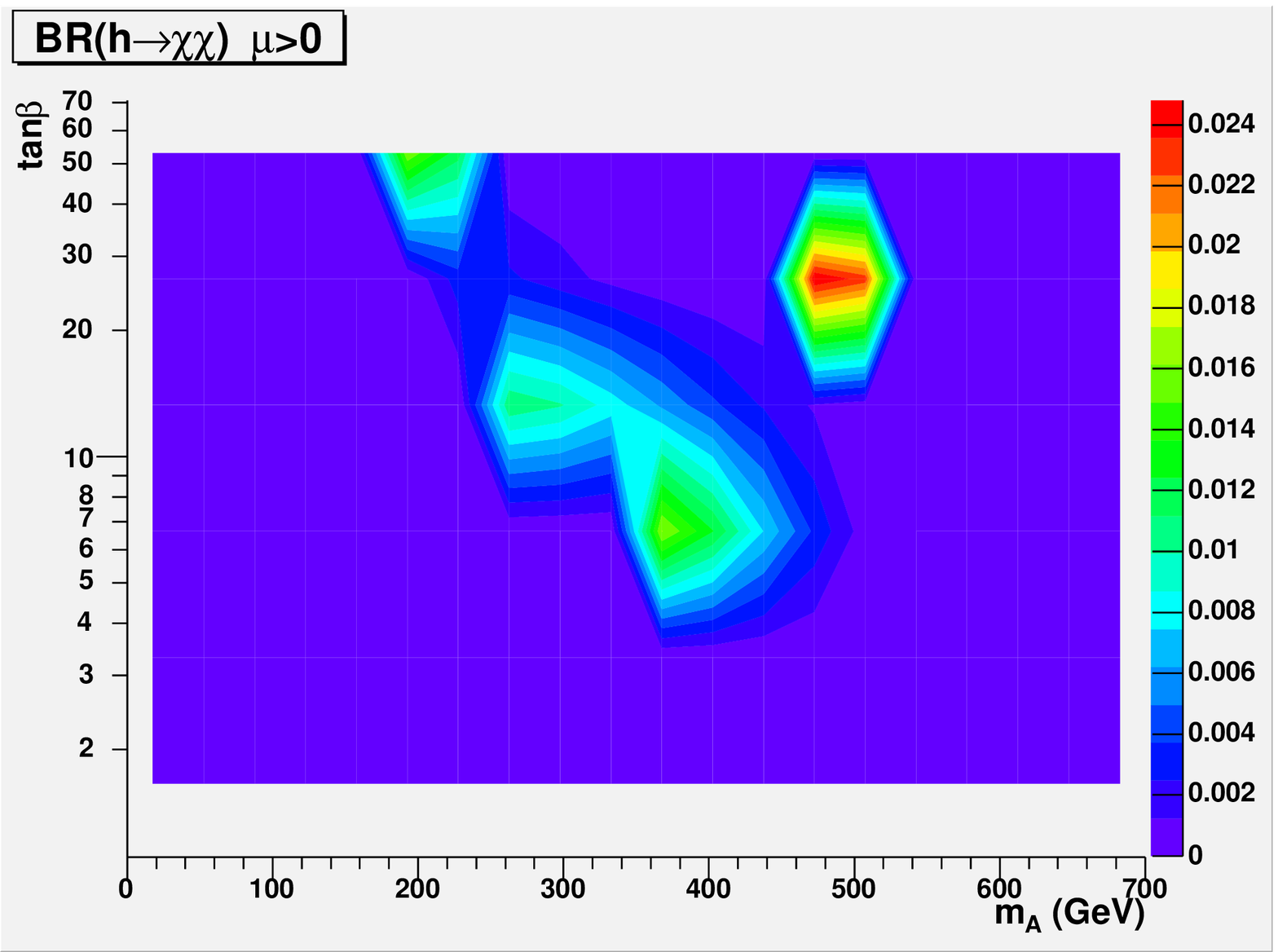}
    \includegraphics[width=0.43\textwidth]{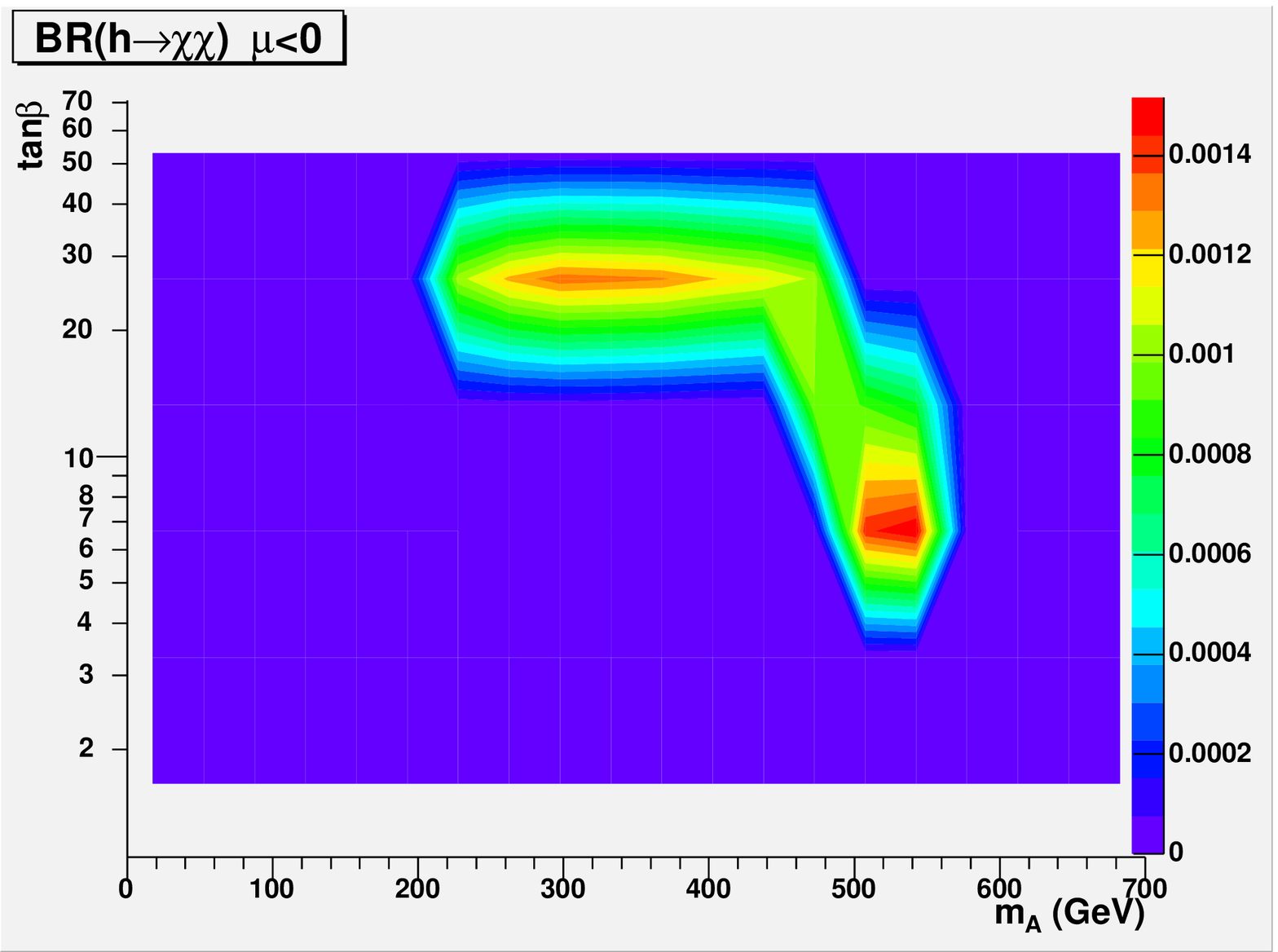}
  \end{center}
  \caption[Scans of mSUGRA model in the $(m_A, tan \beta)$ plane.]{\small The maximal values of $BR(h\to\chi^0_1\chi^0_1)$
in mSUGRA models projected on the $(m_A, tan \beta)$ plane, $A_0=0$, after applying theoretical and experimental constraints.}
 \label{figmachilim}
\end{figure}

\begin{figure}[htb]
  \begin{center}
    \includegraphics[width=0.43\textwidth]{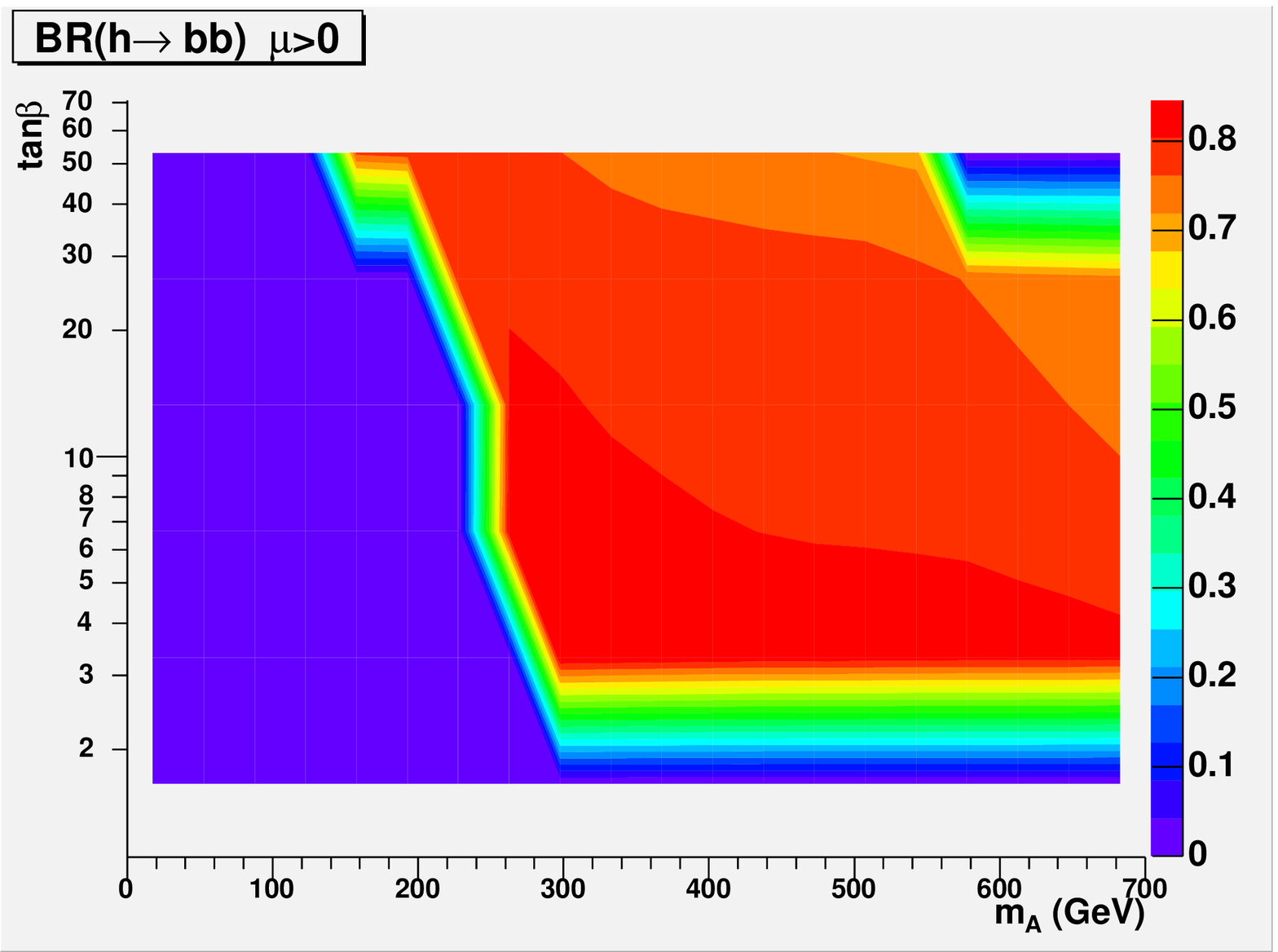}
    \includegraphics[width=0.43\textwidth]{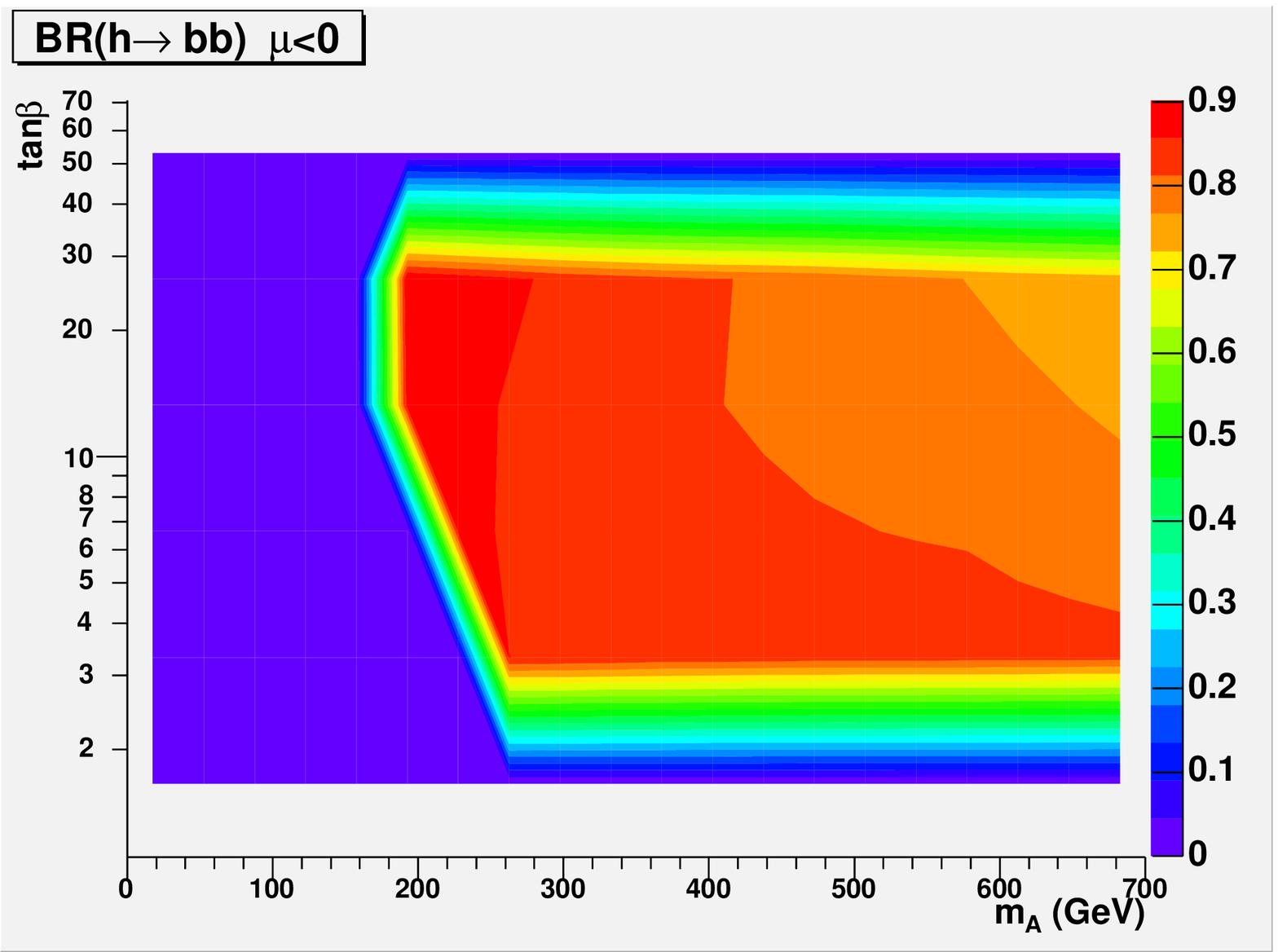}
  \end{center}
  \caption[Scans of mSUGRA model in the $(m_A, tan \beta)$ plane.]{\small The maximum values of $BR(h\to bb)$
in mSUGRA models projected on the $(m_A, tan \beta)$ plane, $A_0=0$, after applying theoretical and experimental constraints.}
 \label{figmabb}
\end{figure}

Figs.~\ref{figmachi}--~\ref{figmabb} show the projection of the mSUGRA model
onto the $(m_A, tan \beta)$ plane. In the mSUGRA model the $m_A$ is 
not given explicitly as an independent variable, but is a function of the initial
parameters. It means that one point from the $(m_A, tan \beta)$ plane may correspond
to more than one $(m_0, m_{1/2})$ pair of initial parameters. In such a case, when plotting the values
of branching ratios, we chose the maximum from these values.

Fig.~\ref{figmachi} shows the branching ratio $BR(h\to\chi^0_1\chi^0_1)$ in a mSUGRA model without 
any theoretical and experimental constraints. We can see that there are regions
where the $\chi^0_1\chi^0_1$ channel can be dominating. However, when we apply the 
constraints according to Table~\ref{T3:limits}, we can see in Fig.~\ref{figmachilim} that all
points with a high branching ratio are excluded, which is also 
seen in the $(m_0, m_{1/2})$ plane. On the other hand, the $BR(h\to b\bar b)$ can be 
very high even with
constraints applied (Fig.~\ref{figmabb}).


\chapter{Conclusions}
\label{chap:conclusions}

In this thesis we studied the prospects for observing the invisibly decaying Higgs boson
in the associated $t\bar tH$ production at the LHC.
The results of Monte Carlo simulations of the signal
and background processes have shown that there is a possibility
of observing a statistically-significant number of signal events
required for the discovery. Moreover, the analysis can still be improved 
to reduce the number of false reconstructions of the $W$ boson.

It is important to note that the analysis of the $t \bar t H$ production described in
Chapter~\ref{chap:analysis} is independent of the model in which the Higgs
boson decays into the invisible channel.
There are several possibilities for models where $H\to invisible$ can be of interest. 
These models include the decay into lightest neutralinos in the supersymmetry models,
or decays into neutrinos in the various models of the neutrino mass generation,
such as extra dimensions, TeV scale gravity, Majorana models or 4th generation lepton.

For this thesis we have studied only the simplest supersymmetric model,
called mSUGRA.
The results of the scans of the mSUGRA model parameter space
show that the regions where the branching ratio of the Higgs particle to the lightest neutralino 
pair is high, are excluded by current experimental constraints. 
The $h \to b \bar b$ channel dominates and the prospects for discovery
in this channel will not be suppressed by opening of invisible decay.
This is very encouraging conclusion:
we will much prefer to have Higgs boson discovered in the visible channel.

In less constrained SUSY models this however might not be the case.
It would
be also interesting to investigate other models that can lead to such a signature,
not just the supersymmetric ones. This is out of the scope of this thesis,
but might provide an interesting research subject.

\end{document}